\documentclass[11pt]{article}
\usepackage{amsfonts}
\usepackage{amssymb}
\usepackage{graphicx}
\usepackage{amsmath}
\usepackage{makeidx}
\usepackage{indentfirst}
\usepackage[T1]{fontenc}
\usepackage[utf8]{inputenc}

\newcounter{resultnum}[section]
\newcounter{conclusionnum}[section]
\newcounter{conditionnum}[section]
\newcounter{conjecturenum}[section]
\newcounter{examplenum}[section]
\newcounter{exercisenum}[section]
\newcounter{lemmanum}[section]
\newcounter{notationnum}[section]
\newcounter{theoremnum}[section]
\newcounter{definitionnum}[section]
\newcounter{corollarynum}[section]
\newcounter{remarknum}[section]
\newcounter{propositionnum}[section]
\newcounter{acknowledgementnum}[section]
\newcounter{algorithmnum}[section]
\newcounter{axiomnum}[section]
\newcounter{casenum}[section]
\newcounter{claimnum}[section]
\newcounter{summarynum}[section]
\newcounter{problemnum}[section]
\begin{document}

\title{Classical and quantum geometric information flows\\
and entanglement of relativistic mechanical systems}
\date{January 30, 2020}
\author{ \vspace{.1 in} \textbf{Sergiu I. Vacaru}
\thanks{
emails: sergiu.vacaru@gmail.com and sergiuvacaru@mail.fresnostate.edu ;\newline
\textit{Address for post correspondence in 2019-2020 as a visitor senior researcher at YF CNU Ukraine:\ } 37 Yu. Gagarin street, ap. 3, Chernivtsi, Ukraine, 58008}  \\
{\small \textit{Physics Department, California State University at Fresno,
Fresno, CA 93740, USA; and }}\\
 {\small \textit{Dep. Theoretical Physics and Computer Modelling, 101  Storozhynetska street,    Chernivtsi, 58029, Ukraine}}
 \vspace{.1 in} \\
\vspace{.1 in} {\ \textbf{Lauren\c{t}iu Bubuianu}}\thanks{%
email: laurentiu.bubuianu@tvr.ro } \\
{\small \textit{SRTV \ - Studioul TVR Ia\c{s}i, \ 28 Alexandru Lapu\c{s}%
neanu street, Ia\c{s}i, \ 700057, Romania;}} \\
{\small and } {\small \textit{University Apollonia, 2 Muzicii street, Ia\c{s}i, \ 700399, Romania }} }
\maketitle

\begin{abstract}
This article elaborates on entanglement entropy and quantum information
theory of geometric flows of (relativistic) Lagrange--Hamilton mechanical
systems. A set of basic geometric and quantum mechanics and probability
concepts together with methods of computation are developed in general
covariant form for curved phase spaces modelled as cotangent Lorentz
bundles. The constructions are based on ideas relating the Grigory
Perelman's entropy for geometric flows and associated statistical
thermodynamic systems to the quantum von Neumann entropy, classical and
quantum relative and conditional entropy, mutual information etc. We
formulate the concept of the entanglement entropy of quantum geometric
information flows and study properties and inequalities for quantum,
thermodynamic and geometric entropies characterising such systems.

\vskip3pt

\textbf{Keywords:}\ Perelman W-entropy; quantum geometric information flows.

\vskip3pt

PACS2010:\ 02.40.-k, 02.90.+p, 03.65.Ud,  04.50.-h, 04.90.+e, 05.90.+m
05.90.+m

MSC2010:\ 53C44, 53C50, 53C80, 81P45, 82D99, 83C15, 83C55, 83C99, 83D99, 35Q75,
37J60, 37D35
\end{abstract}

\tableofcontents

\section{Introduction}

A generic feature of quantum physics which is absent in classical physics is
that of entanglement. There were introduced several entanglement measures of
how much quantum a given system is. Because of computational accessibility,
the entanglement entropy plays a particulary important role together with R%
\'{e}nyi entropies, mutual information etc. For recent reviews of most
important ideas and results related to quantum information theory, we cite \
\cite%
{preskill,solodukhin11,aolita14,ionicioiu16,stoica18,nishioka18,witten18,witten18a,ecker18}
and references therein. Here we note that the concept of entanglement
entropy originated from quantum information theory \cite{nielsen10}. At
present, it is connected to a wide range of applications in condensed matter
physics, gravity theories and particle physics etc. The progress in such
directions included the holographic formula for entanglement entropy \cite%
{ryu06}, a new type of order parameter for quantum-phase transitions \cite%
{vidal02,kitaev05,fendley06}, ideas of formulating quantum gravity from
quantum entanglement and so-called ER=EPR \cite{raamsdonk10,maldacena13}.

There are many motivations to study quantum entanglement which depends on
respective directions of research. For instance, we elaborated \cite%
{vacaru19,vacaru19a} on the idea that an intriguing connection exists
between the Poincar\'{e}-Thurston conjecture (it became again a conjecture
for relativistic Ricci flows even a proof exists for Riemannian metrics \cite%
{perelman1}) and the emergent entropic gravity and/or other type
modifications.  G. Perelman introduced and applied in his famous preprints \cite{perelman1,perelman2,perelman3} the F- and W-functionals from which the R. Hamilton's Ricci flow equations for Ricci flows \cite{hamilt1,hamilt2,hamilt3} can be derived. Here we note that in physics such equations were considered earlier by D. Friedan \cite{friedan1,friedan2,friedan3}.  In a general context, such works and further developments  provide strong motivations
for elaborating a new direction (based on geometric flows and associated
thermodynamical models) in classical and quantum information theory. For such
models, the quantum entanglement can be exploited for computational tasks
which are impossible if only classical methods are used but for performing on
new type theories unifying quantum and geometric flow evolution scenarios.

This article is the 5th partner in a series of works \cite{bubuianu19,vacaru19,vacaru19a,vacaru19b} devoted to applications of G. Perelman's entropic functionals \cite{perelman1} and nonholonommic geometric
flow methods in classical and quantum information theory, geometric mechanics and thermodynamics, and modified (entropic and other types) gravity. For a review of mathematical results on Ricci flows of Riemannian and K\"{a}hler metrics, rigorous proofs and topological and geometric analysis methods, we cite \cite{monogrrf1,monogrrf2,monogrrf3}. In our approach, we consider nonholonomic deformations of the G. Perelman's functionals and elaborated on new geometric methods and applications in (modified) gravity, geometric mechanics; locally anisotropic kinetics, diffusion and thermodynamics; and information theory. Here we note that in this work we follow the notations on the so-called quantum geometric information flow, QGIF, theory (in brief, it is used GIF for classical models) introduced in \cite{vacaru19b}. Readers may study our previous works \cite{vacaru00ap,vacaru09,rajpoot17,ruchin13,gheorghiu16} and references therein,
on nonholonomic (non) commutative / supersymmetric geometric flows and
related kinetic and statistical thermodynamic models.

The aim of this paper is to specifically address the geometric flow evolution and dynamics of the entanglement in quantized Lagrange--Hamilton relativistic mechanical systems. We develop our approach on elaborating new principles and methods for formulating classical and quantum information theories encoding geometric flows and their analogous geometric thermodynamic models. The key ideas for developing such
new directions in (quantum) information theory and applications is to extend the standard constructions involving the von Neumann, and related conditional and relative entropies. We introduce into consideration
generalizations of the concepts of W-entropy and analogous thermodynamic entropy elaborated in original variants by G. Perelman for Ricci flows of Riemannian metrics.

We try to make this work self-contained and multi-disciplinary pedagogic enough but for advanced researchers working on geometry and physics, nonholonomic geometric mechanics and thermodynamics, quantum mechanics and quantum field theory and information theory. In our case, some typical Alice
and Bob communicating using methods of quantum information theory should also have certain knowledge on geometric flows; systems of nonlinear partial differential equations, PDEs, and their applications in modern classical and quantum physics. It is assumed that readers have a background on modified gravity theories and modern astrophysics and cosmology because all such theories provide strong motivations and examples of applications of the formalism elaborated in the cited monographs, reviews and series of works on
geometric flows and information theory. In this article, we study entanglement for quantum geometric flows of mechanical systems and do not concern issues on gravity, quantum field theory or condensed matter physics.
On emergent gravity theories, modified Ricci flow theories and gravity, exact solutions and related classical and quantum mechanical entropic functionals from which generalized Einstein equations can be derived, see
our recent results \cite{vacaru00ap,vacaru09,rajpoot17,ruchin13,gheorghiu16} and references therein.

This article is organized as follows: In section \ref{s2}, we start with reviewing the fundamentals of the theory of geometric flows of relativistic Hamilton phase spaces. After defining the fundamental geometric objects such as the nonlinear connection, N-connection, and distinguished metric, d-metric, structures, we show how the curvatures can be computed for general and preferred linear connections. Then we introduce the G.
Perelman F- and W-functionals (entropic type) for W. Hamilton mechanical systems and their formulation in general N-adapted variables.

Section \ref{s3} begins with a quick introduction into the statistical thermodynamic theory of geometric information flows, GIFs, when the G. Perelman approach is generalized for nonholonomic N-adapted variables. The approach is generalized for quantum geometric information flows, QGIFs, using the statistical density matrix and its analogous quantum density matrix. The von Neumann entropy for QGIFs and quantum generalizations of the W- and thermodynamic entropy are considered.

In section \ref{s4}, we explore the entanglement and QGIFs as quantum mechanical systems. There are defined QGIF analogs of two spin systems, thermofield double GIF states and Bell like geometric flow states. We outline the main properties and inequalities of the entanglement entropy for such systems with mixed geometric and quantum flow evolution. The entanglement and R\'{e}nyi entropy and QGIFs at finite temperature are studied.  Conclusions are provided in section \ref{s5}.

\section{Geometric flows of relativistic Hamilton phase spaces}

\label{s2} We present a short review of the geometry of relativistic
Hamilton phase spaces modelled on cotangent bundle $T^{\ast }\mathbf{V}$ of
a nonholonomic Lorentz manifold $\mathbf{V}$, see an axiomatic approach and
details in \cite{vacaru18,bubuianu18}. There are provided formulas for
respective generalizations of G. Perelman's F- and W-entropy functionals for
which we follow the conventions from \cite{vacaru19,vacaru19a,vacaru19b},
see proofs and references therein.

\subsection{The Hessian geometry of relativistic Hamilton spaces}

\subsubsection{Nonlinear connections and adapted metrics}

We consider a cotangent Lorentz bundle $T^{\ast }V,\dim V=4,$ enabled with
local coordinates$\ ^{\shortmid }u^{\alpha }=(x^{i},p_{a}),$ (in brief, $\
^{\shortmid }u=(x,p)),$ where $x^{i}$ are base manifold coordinates and $%
p_{a}$ are momentum like typical fiber coordinates. Such a model of
relativistic phase spacetime is enabled in any point with a total metric
structure (phase space metric) of signature $(+++-;+++-),$ which for
corresponding frames/coordinates transforms can be represented in the form
\begin{equation}
d\ ^{\shortmid }s^{2}=\ ^{\shortmid }g_{\alpha \beta }d\ ^{\shortmid
}u^{\alpha }d\ ^{\shortmid }u^{\beta }=\eta _{ij}dx^{i}dx^{j}+\eta
^{ab}dp_{a}dp_{b},\mbox{ for }p_{a}\sim dx_{a}/d\tau .  \label{lqed}
\end{equation}%
In these formulas, $\eta _{ij}=diag[1,1,1,-1]$ and $\eta
^{ab}=diag[1,1,1,-1].$\footnote{\label{fncoordconv}We follow such
conventions:\ the "horizontal" indices, h--indices, run values $%
i,j,k,...=1,2,3,4;$ the vertical indices, v-vertical, run values $%
a,b,c...=5,6,7,8$; respectively, the v-indices can be identified/ contracted
with h-indices $1,2,3,4$ for lifts on total (co) bundles, when $\alpha
=(i,a),\beta =(j,b),\gamma =(k,c),...=1,2,3,...8.$ There are used letters
labelled by an abstract left up/low symbol "$\ ^{\shortmid }$" (for
instance, $\ ^{\shortmid }u^{\alpha }$ and $\ ^{\shortmid }g_{\alpha \beta
}) $ in order to emphasize that certain geometric/ physical objects are
defined on $T^{\ast }V.$ Similar formulas can be derived on $TV$ for
geometric objects labeled without "$\ ^{\shortmid }$". Boldface symbols are
used for geometric objects on spaces endowed with nonlinear connection
structure (see below formula (\ref{nconcan})).} In a more general context,
we can elaborate on physical models on curved phase spaces when the metric
structure (\ref{lqed}) is determined by coefficients of type $\ ^{\shortmid
}g_{\alpha \beta }(\ ^{\shortmid }u)=[g_{ij}(x),\ ^{\shortmid }g^{ab}(x,p)].$

A relativistic Hamilton space $H^{3,1}=(T^{\ast }V,H(x,p))$ is determined by
a fundamental function $H(x,p)$ (it can be used a generating
Hamilton function, Hamiltonian or Hamilton density). For classical models,
it is considered that a map $T^{\ast }V\ni (x,p)\rightarrow H(x,p)\in
\mathbb{R}$ defines a real valued function being differentiable on $%
\widetilde{T^{\ast }V}:=T^{\ast }V/\{0^{\ast }\}$, for $\{0^{\ast }\}$ being
the null section of $T^{\ast }V,$ and continuous on the null section of $\pi
^{\ast }:\ T^{\ast }V\rightarrow V.$ In a more general context, a $H(x,p)$
can be quantized following prescriptions for a respective quantum model
(quantum mechanics, QM, or quantum field theory, QFT, with corresponding
quasi-classical relativistic and non-relativistic limits). In this work, we
elaborate on relativistic mechanical models which are regular if the Hessian
(cv-metric)
\begin{equation}
\ ^{\shortmid }\widetilde{g}^{ab}(x,p):=\frac{1}{2}\frac{\partial ^{2}H}{%
\partial p_{a}\partial p_{b}}  \label{hesshs}
\end{equation}%
for a $H=\widetilde{H}$ is non-degenerate, i.e. $\det |\ ^{\shortmid }%
\widetilde{g}^{ab}|\neq 0,$ and of constant signature.

For Lagrange and Hamilton spaces, we can perform Legendre transforms $%
L\rightarrow H(x,p):=p_{a}y^{a}-L(x,y)$ and $y^{a}$ determining solutions of
the equations $p_{a}=\partial L(x,y)/\partial y^{a}.$ In a similar manner,
the inverse Legendre transforms can be introduced, $H\rightarrow L,$ when $%
L(x,y):=p_{a}y^{a}-H(x,p)$ for $p_{a}$ determining solutions of the
equations $y^{a}=\partial H(x,p)/\partial p_{a}.$ In this work, we consider
Hamilton structures which allow canonical Hamilton formulations of some QM
models and respective quasi-classical limits.

Any $\widetilde{H}$ defines a canonical nonlinear connection (N-connection) structure
\begin{equation}
\ ^{\shortmid }\widetilde{\mathbf{N}}:\ TT^{\ast }V=hT^{\ast }V\oplus
vT^{\ast }V  \label{nconcan}
\end{equation}%
and a N-adapted canonical distinguished metric (d-metric) structure
parameterized with conventional horizontal, h, and covertical, cv,
components,%
\begin{equation}
\ ^{\shortmid }\widetilde{\mathbf{g}}=\ ^{\shortmid }\widetilde{\mathbf{g}}%
_{\alpha \beta }(x,p)\ ^{\shortmid }\widetilde{\mathbf{e}}^{\alpha }\mathbf{%
\otimes \ ^{\shortmid }}\widetilde{\mathbf{e}}^{\beta }=\ \ ^{\shortmid }%
\widetilde{g}_{ij}(x,p)e^{i}\otimes e^{j}+\ ^{\shortmid }\widetilde{g}%
^{ab}(x,p)\ ^{\shortmid }\widetilde{\mathbf{e}}_{a}\otimes \ ^{\shortmid }%
\widetilde{\mathbf{e}}_{b},  \label{cdmds}
\end{equation}%
where the canonical N-linear frames $\ ^{\shortmid }\widetilde{\mathbf{e}}%
^{\alpha }=(e^{i},\ ^{\shortmid }\widetilde{\mathbf{e}}_{a})$ are
canonically determined by data $(\widetilde{H},\ ^{\shortmid }\widetilde{g}%
^{ab}).$\footnote{%
The coefficients of the canonical N-connection are computed following
formulas \newline
$\ ^{\shortmid }\widetilde{\mathbf{N}}=\left\{ \ ^{\shortmid }\widetilde{N}%
_{ij}:=\frac{1}{2}\left[ \{\ \ ^{\shortmid }\widetilde{g}_{ij},\widetilde{H}%
\}-\frac{\partial ^{2}\widetilde{H}}{\partial p_{k}\partial x^{i}}\
^{\shortmid }\widetilde{g}_{jk}-\frac{\partial ^{2}\widetilde{H}}{\partial
p_{k}\partial x^{j}}\ ^{\shortmid }\widetilde{g}_{ik}\right] \right\}$,
where $\ \ ^{\shortmid }\widetilde{g}_{ij}$ is inverse to $\ \ ^{\shortmid }%
\widetilde{g}^{ab}$ (\ref{hesshs}). The canonical N--adapted (co) frames are
\begin{equation*}
\ ^{\shortmid }\widetilde{\mathbf{e}}_{\alpha }=(\ ^{\shortmid }\widetilde{%
\mathbf{e}}_{i}=\frac{\partial }{\partial x^{i}}-\ ^{\shortmid }\widetilde{N}%
_{ia}(x,p)\frac{\partial }{\partial p_{a}},\ ^{\shortmid }e^{b}=\frac{%
\partial }{\partial p_{b}});\ \ ^{\shortmid }\widetilde{\mathbf{e}}^{\alpha
}=(\ ^{\shortmid }e^{i}=dx^{i},\ ^{\shortmid }\mathbf{e}_{a}=dp_{a}+\
^{\shortmid }\widetilde{N}_{ia}(x,p)dx^{i}),
\end{equation*}%
being characterized by corresponding anholonomy relations $\ [\ ^{\shortmid }%
\widetilde{\mathbf{e}}_{\alpha },\ ^{\shortmid }\widetilde{\mathbf{e}}%
_{\beta }]=\ ^{\shortmid }\widetilde{\mathbf{e}}_{\alpha }\ ^{\shortmid }%
\widetilde{\mathbf{e}}_{\beta }-\ ^{\shortmid }\widetilde{\mathbf{e}}_{\beta
}\ ^{\shortmid }\widetilde{\mathbf{e}}_{\alpha }=\ ^{\shortmid }\widetilde{W}%
_{\alpha \beta }^{\gamma }\ ^{\shortmid }\widetilde{\mathbf{e}}_{\gamma },$
with anholonomy coefficients $\widetilde{W}_{ia}^{b}=\partial _{a}\widetilde{%
N}_{i}^{b},$ $\widetilde{W}_{ji}^{a}=\widetilde{\Omega }_{ij}^{a},$ and $\
^{\shortmid }\widetilde{W}_{ib}^{a}=\partial \ ^{\shortmid }\widetilde{N}%
_{ib}/\partial p_{a}$ and $\ ^{\shortmid }\widetilde{W}_{jia}=\ \mathbf{\
^{\shortmid }}\widetilde{\Omega }_{ija}.$ Such a frame is holonomic
(integrable) if the respective anholonomy coefficients are zero.}

Considering general frame (vierbein) transforms, $e_{\alpha }=e_{\ \alpha }^{%
\underline{\alpha }}(u)\partial /\partial u^{\underline{\alpha }}$ and $%
e^{\beta }=e_{\ \underline{\beta }}^{\beta }(u)du^{\underline{\beta }},$ any
N-connection and d-metric structure on a cotangent Lorentz bundle $T^{\ast }%
\mathbf{V}$ can be written in general form (without "tilde" on symbols),%
\begin{eqnarray*}
\ ^{\shortmid }\mathbf{N} &=&\{\ ^{\shortmid }N_{ij}(x,p)\},%
\mbox{ with
arbitrary coefficients }; \\
\ ^{\shortmid }\mathbf{g} &=&\ ^{\shortmid }\mathbf{g}_{\alpha \beta }(x,p)\
^{\shortmid }\mathbf{e}^{\alpha }\mathbf{\otimes \ ^{\shortmid }e}^{\beta
}=\ \ ^{\shortmid }g_{ij}(x,p)e^{i}\otimes e^{j}+\ ^{\shortmid }g^{ab}(x,p)\
^{\shortmid }\mathbf{e}_{a}\otimes \ ^{\shortmid }\mathbf{e}_{b}.
\end{eqnarray*}%
So, any classical regular Hamilton mechanics can be geometrized in general
form on a phase spacetime $T^{\ast }\mathbf{V}$ by some nonholonomic data $(\
^{\shortmid }\mathbf{N,}\ ^{\shortmid }\mathbf{g}).$ Inversely, using
respective frame transforms on a nonholonomic cotangent bundle, we can
always consider a relativistic Hamilton space model defined by some data $(%
\widetilde{H},\ ^{\shortmid }\widetilde{\mathbf{N}};\ ^{\shortmid }%
\widetilde{\mathbf{e}}_{\alpha },\ ^{\shortmid }\widetilde{\mathbf{e}}%
^{\alpha };\ \ ^{\shortmid }\widetilde{g}^{ab},\ \ ^{\shortmid }\widetilde{g}%
_{ab}).$

\subsubsection{Curvatures, torsions and nonmetricity of linear and
distinguished connections}

A physically realistic geometrization of physical models on $T^{\ast }%
\mathbf{V}$ is possible if such a phase space is enabled with a linear
(affine) connection structure. Using $\ ^{\shortmid }\mathbf{g}$, we can
define in standard form the Levi-Civita connection $\ ^{\shortmid }\nabla $
(as a unique metric compatible and with zero torsion) but such a geometric
object is not adapted to the N-connection structure. To elaborate N-adapted
geometric models we have to consider the concept of distinguished connection
(d--connection) which is a linear connection $\ ^{\shortmid }\mathbf{D}$ on $%
\mathbf{T}^{\ast }\mathbf{V}$ preserving under parallel transports a
N--connection splitting $\ ^{\shortmid }\mathbf{N}$. With respect to a
N-adapted basis the coefficients of a d-connection $\ ^{\shortmid }\mathbf{D}
$ are labelled $\ ^{\shortmid }\mathbf{\Gamma }_{\ \beta \gamma }^{\alpha
}=\{\ ^{\shortmid }L_{\ jk}^{i},\ ^{\shortmid }\acute{L}_{a\ k}^{\ b},\
^{\shortmid }\acute{C}_{\ j}^{i\ c},\ ^{\shortmid }C_{a}^{\ bc}\}$. This
involves an explicit h-- and cv--splitting, of covariant derivatives $\
^{\shortmid }\mathbf{D}=\left( \ _{h}^{\shortmid }\mathbf{D,\ }%
_{cv}^{\shortmid }\mathbf{D}\right) ,$ where $\ _{h}^{\shortmid }\mathbf{D}%
=\{\ ^{\shortmid }L_{\ jk}^{i},\ ^{\shortmid }\acute{L}_{a\ k}^{\ b}\}$ and $%
\ _{cv}^{\shortmid }\mathbf{D}=\{\ ^{\shortmid }\acute{C}_{\ j}^{i\ c},\
^{\shortmid }C_{a}^{\ bc}\}.$

Prescribing a d-connection structure $\ ^{\shortmid }\mathbf{D,}$ we can
work alternatively with an arbitrary linear connection a linear connection $%
\underline{D}$ (which is not obligatory a d-connection) $\ ^{\shortmid }%
\underline{D}$ on $\mathbf{T}^{\ast }\mathbf{V.}$ For such covector bundles,
there are nonholonomic deformation relations with a respective distortion
distinguished tensor, d-tensor, $\ ^{\shortmid }\mathbf{Z:=\ ^{\shortmid }D}%
-\ ^{\shortmid }\underline{D}.$

For any linear connection and/or d-connection structure, $\ ^{\shortmid }%
\underline{D}$ and/or $\ ^{\shortmid }\mathbf{D}$, we can define in standard
form respective curvature, $\ ^{\shortmid }$\underline{$\mathcal{R}$} and/or
$\ ^{\shortmid }\mathcal{R},$ torsion, $\ ^{\shortmid }\underline{\mathcal{T}%
}$ and/or $\ ^{\shortmid }\mathcal{T},$ nonmetricity, $\ \ ^{\shortmid }%
\underline{\mathcal{Q}}$ and/or $\ ^{\shortmid }\mathcal{Q},$ d-tensors,
\begin{eqnarray}
\ ^{\shortmid }\underline{\mathcal{R}}(\ ^{\shortmid }\mathbf{X,\
^{\shortmid }Y}) &:=&\ ^{\shortmid }\underline{D}_{\ ^{\shortmid }\mathbf{X}%
}\ \ ^{\shortmid }\underline{D}_{\ ^{\shortmid }\mathbf{Y}}-\ ^{\shortmid }%
\underline{D}_{\ ^{\shortmid }\mathbf{Y}}\ \ ^{\shortmid }\underline{D}_{\
^{\shortmid }\mathbf{X}}-\ \ ^{\shortmid }\underline{D}_{\mathbf{[\
^{\shortmid }X,\ ^{\shortmid }Y]}},  \label{dcurvabstr} \\
\ ^{\shortmid }\underline{\mathcal{T}}(\ ^{\shortmid }\mathbf{X,\
^{\shortmid }Y}) &:=&\ \ ^{\shortmid }\underline{D}_{\ ^{\shortmid }\mathbf{X%
}}\ ^{\shortmid }\mathbf{Y}-\ \ ^{\shortmid }\underline{D}_{\ ^{\shortmid }%
\mathbf{Y}}\ ^{\shortmid }\mathbf{X}-[\ ^{\shortmid }\mathbf{X,\ ^{\shortmid
}Y}],\ \ ^{\shortmid }\underline{\mathcal{Q}}(\ ^{\shortmid }\mathbf{X}):=\
^{\shortmid }\underline{D}_{\ ^{\shortmid }\mathbf{X}}\ ^{\shortmid }\mathbf{%
g}, \mbox{ and/or }  \notag \\
\ ^{\shortmid }\mathcal{R}(\ ^{\shortmid }\mathbf{X,\ ^{\shortmid }Y})&:=& \
^{\shortmid }\mathbf{D}_{\ ^{\shortmid }\mathbf{X}}\ ^{\shortmid }\mathbf{D}%
_{\ ^{\shortmid }\mathbf{Y}}-\ ^{\shortmid }\mathbf{D}_{\ ^{\shortmid }%
\mathbf{Y}}\ ^{\shortmid }\mathbf{D}_{\ ^{\shortmid }\mathbf{X}}-\
^{\shortmid }\mathbf{D}_{\mathbf{[\ ^{\shortmid }X,\ ^{\shortmid }Y]}},
\notag \\
\ ^{\shortmid }\mathcal{T}(\ ^{\shortmid }\mathbf{X,\ ^{\shortmid }Y}) &:=&\
^{\shortmid }\mathbf{D}_{\ ^{\shortmid }\mathbf{X}}\ ^{\shortmid }\mathbf{Y}%
-\ ^{\shortmid }\mathbf{D}_{\ ^{\shortmid }\mathbf{Y}}\ ^{\shortmid }\mathbf{%
X}-[\ ^{\shortmid }\mathbf{X,\ ^{\shortmid }Y}],\ ^{\shortmid }\mathcal{Q}(\
^{\shortmid }\mathbf{X}):=\ ^{\shortmid }\mathbf{D}_{\ ^{\shortmid }\mathbf{X%
}}\ ^{\shortmid }\mathbf{g.}  \notag
\end{eqnarray}
The N--adapted and/or coordinate formulas for coefficients of such geometric
objects can be computed in explicit form, see appendices to \cite%
{vacaru18,bubuianu18} and references therein.

Using (\ref{dcurvabstr}), we can define and compute respective Ricci
tensors/ d-tensors, scalar curvatures etc. For instance, the Ricci d--tensor
of a $^{\shortmid }\mathbf{D}$ is defined and computed $\ ^{\shortmid
}Ric=\{\ ^{\shortmid }\mathbf{R}_{\alpha \beta }:=\ ^{\shortmid }\mathbf{R}%
_{\ \alpha \beta \tau }^{\tau }\}.$ The N-adapted coefficients of the Ricci
d--tensor of a d-connection $\ ^{\shortmid }\mathbf{D}$ in respective phase
spaces are parameterized in $h$- and/or $cv$-form by formulas
\begin{equation}
\ ^{\shortmid }\mathbf{R}_{\alpha \beta }=\{\ ^{\shortmid }R_{hj}:=\
^{\shortmid }R_{\ hji}^{i},\ ^{\shortmid }R_{j}^{\ a}:=-\ ^{\shortmid }P_{\
ji}^{i\ \ \ a},\ \ ^{\shortmid }R_{\ k}^{b}:=\ ^{\shortmid }P_{a\ k}^{\ b\ \
a},\ ^{\shortmid }R_{\ }^{bc}=\ ^{\shortmid }S_{a\ }^{\ bca}\}.
\label{driccid}
\end{equation}%
Such formulas for $\ ^{\shortmid }\underline{D}$ can be written in a similar
"underlined" form. Hereafter, for simplicity, we shall provide the formulas
only for a general d-connection $\ ^{\shortmid }\mathbf{D}$ if that will not
result in ambiguities.

If a phase space is enabled both with a d-connection, $\ ^{\shortmid }%
\mathbf{D,}$ and a d-metric, $\ ^{\shortmid }\mathbf{g,}$ structures, we can
define and compute nonholonomic Ricci scalars,
\begin{equation}
\ _{s}^{\shortmid }R:=\ ^{\shortmid }\mathbf{g}^{\alpha \beta }\ ^{\shortmid
}\mathbf{R}_{\alpha \beta }=\ ^{\shortmid }g^{ij}\ ^{\shortmid }R_{ij}+\
^{\shortmid }g^{ab}\ ^{\shortmid }R_{ab}=\ ^{\shortmid }R+\ ^{\shortmid }S,
\label{dricciscal}
\end{equation}%
with respective h-- and v--components, $\ ^{\shortmid }R=\ ^{\shortmid
}g^{ij}\ ^{\shortmid }R_{ij}$ and$\ ^{\shortmid }S=\ ^{\shortmid }g_{ab}\
^{\shortmid }S^{ab}.$

The geometric objects (\ref{dcurvabstr}), (\ref{driccid}) and (\ref%
{dricciscal}) can be defined for any special classes of linear connection
structures. In next subsection, we consider three important classes of
linear and/or d-connections determined by a d-metric structure $\
^{\shortmid }\widetilde{\mathbf{g}}$ or $\ ^{\shortmid }\mathbf{g}$.

\subsubsection{Preferred linear and d-connection structures}

Any relativistic phase space $\mathbf{T}^{\ast }\mathbf{V}$ can be described
as a Hamilton space using the canonical data $(\ ^{\shortmid }\widetilde{%
\mathbf{N}},\ ^{\shortmid }\widetilde{\mathbf{g}}) $ and/or in general
nonholonomic (pseudo) Riemannian form for some $\left( \ ^{\shortmid }%
\mathbf{N},\ ^{\shortmid }\mathbf{g}\right)$. Respective canonical
N--connections $\ ^{\shortmid }\widetilde{\mathbf{N}}$ and/or $\ ^{\shortmid
}\mathbf{N}$ define correspondingly certain canonical almost complex
structures $\ ^{\shortmid }\widetilde{\mathbf{J}}$ and/or $\ ^{\shortmid }%
\mathbf{J}.$ For instance, we can consider a linear operator $\ ^{\shortmid }%
\mathbf{J}$ acting on$\ ^{\shortmid }\mathbf{e}_{\alpha }=(\ ^{\shortmid }%
\mathbf{e}_{i},\ ^{\shortmid }e^{b})$ using formulas $\ ^{\shortmid }\mathbf{%
J}(\ ^{\shortmid }\mathbf{e}_{i})=-\ ^{\shortmid }e^{n+i}$ and $\
^{\shortmid }\mathbf{J}(\ ^{\shortmid }e^{n+i})=\ ^{\shortmid }\mathbf{e}%
_{i} $. Such a $\ ^{\shortmid }\mathbf{J}$ defines globally an almost
complex structure ( $\ \ ^{\shortmid }\mathbf{J\circ \ }\ ^{\shortmid }%
\mathbf{J}=$ $-\mathbf{\ I,}$ where $\mathbf{I}$ is the unity matrix) on $%
\mathbf{T}^{\ast }\mathbf{V.}$ Using $\ ^{\shortmid }\widetilde{\mathbf{J}}$
and $\ ^{\shortmid }\mathbf{J,}$ we can define respective (canononical)
almost symplectic structures, $\ ^{\shortmid }\widetilde{\theta }:=\
^{\shortmid }\widetilde{\mathbf{g}}(\ ^{\shortmid }\widetilde{\mathbf{J}}%
\cdot ,\cdot )$ and $\ ^{\shortmid }\theta :=\ ^{\shortmid }\mathbf{g}(\
^{\shortmid }\mathbf{J}\cdot ,\cdot ).$ In result, we can construct such
preferred linear/distinguished connections:
\begin{equation}
\begin{array}{ccccc}
\ ^{\shortmid }\nabla : &  & \ ^{\shortmid }\nabla \ ^{\shortmid }\mathbf{g}%
=0;\ \ ^{\shortmid }\mathbf{T[\ ^{\shortmid }\nabla ]}=0, &  & %
\mbox{Hamilton LC-connection}; \\
\ ^{\shortmid }\widehat{\mathbf{D}}: &  & \ ^{\shortmid }\widehat{\mathbf{D}}%
\ \mathbf{g}=0;\ h\ ^{\shortmid }\widehat{\mathbf{T}}=0,\ cv\ ^{\shortmid }%
\widehat{\mathbf{T}}=0. &  & \mbox{canonical Hamilton d-connection}; \\
\ ^{\shortmid }\widetilde{\mathbf{D}}: &  & \ ^{\shortmid }\widetilde{%
\mathbf{D}}\ ^{\shortmid }\widetilde{\theta }=0,\ ^{\shortmid }\widetilde{%
\mathbf{D}}\ ^{\shortmid }\widetilde{\theta }=0 &  &
\mbox{almost sympl.
Hamilton d-connection.}%
\end{array}
\label{prefercon}
\end{equation}

The geometric objects in (\ref{prefercon}) are related via corresponding
distortion relations
\begin{equation}
\ ^{\shortmid }\widehat{\mathbf{D}}=\ ^{\shortmid }\nabla +\ ^{\shortmid }%
\widehat{\mathbf{Z}},\ ^{\shortmid }\widetilde{\mathbf{D}} =\
^{\shortmid}\nabla +\ ^{\shortmid }\widetilde{\mathbf{Z}}, \mbox{ and } \
^{\shortmid }\widehat{\mathbf{D}}=\ ^{\shortmid }\widetilde{\mathbf{D}} + \
^{\shortmid}\mathbf{Z}, \mbox{ determined by } (\ ^{\shortmid }\mathbf{g,\
^{\shortmid }N)};  \notag
\end{equation}%
with distortion d-tensors$\ ^{\shortmid }\widehat{\mathbf{Z}},\ ^{\shortmid }%
\widetilde{\mathbf{Z}},$ and $\ ^{\shortmid }\mathbf{Z}$, on $T\mathbf{T}%
^{\ast }\mathbf{V}$. In principle, we can work with any such linear
connection structure even they have different geometric and physical
meaning. The corresponding curvatures and Ricci d-tensors and scalar
curvatures can be computed by introducing such distortion relations in
respective formulas (\ref{dcurvabstr}), (\ref{driccid}) and (\ref{dricciscal}).

\subsection{F- and W-functionals for mechanical systems in general N-adapted
variables}

The goal of this subsection is to generalize G. Perelman's functionals and
formulate and approach to the theory of nonholonomic geometric flows of
relativistic mechanical systems. We shall consider canonical Hamilton
variables and nonholonomic deformations to a general d-connection structure.
This is important for further developments in classical and quantum
information theories when the Hamilton variables are used in explicit form
for analyzing certain analogous mechanical and thermodynamic models and,
latter, the results are reformulated in general covariant forms.

We consider a family of nonholonomic cotangent Lorentz bundles $T^{\ast }%
\mathbf{V}(\tau )$ enabled with corresponding sets of  canonical
N--connections $\ ^{\shortmid }\widetilde{\mathbf{N}}(\tau )=\ ^{\shortmid }%
\widetilde{\mathbf{N}}(\tau ,\ ^{\shortmid }u)$ and d-metrics $\ ^{\shortmid
}\widetilde{\mathbf{g}}(\tau )=\ ^{\shortmid }\widetilde{\mathbf{g}}(\tau ,\
^{\shortmid }u)$ all parameterized by a positive parameter $\tau ,0\leq \tau
\leq \tau _{0}.$\footnote{%
For simplicity, we shall write, for instance, $\ ^{\shortmid }\widetilde{%
\mathbf{N}}(\tau )$ instead of $\ ^{\shortmid }\widetilde{\mathbf{N}}(\tau
,\ ^{\shortmid }u)$ if that will not result in ambiguities. Relativistic
nonholonomic phase spacetimes can be enabled with necessary types double
nonholonomic (2+2)+(2+2) \ and (3+1)+(3+1) splitting \cite%
{rajpoot17,ruchin13,gheorghiu16,bubuianu19,vacaru19}. Local (3+1)+(3+1)
coordinates are labeled in the form $\ ^{\shortmid }u=\{\ ^{\shortmid
}u^{\alpha }=\ ^{\shortmid }u^{\alpha
_{s}}=(x^{i_{1}},y^{a_{2}};p_{a_{3}},p_{a_{4}})=(x^{\grave{\imath}%
},u^{4}=y^{4}=t;p_{\grave{a}},p_{8}=E)\}$ for $i_{1},j_{1},k_{1},...=1,2;$ $%
a_{1},b_{1},c_{1},...=3,4;$ $a_{2},b_{2},c_{2},...=5,6;$ $%
a_{3},b_{3},c_{3},...=7,8.$ The insices $\grave{\imath},\grave{j},\grave{k}%
,...=1,2,3,$ respectively, $\grave{a},\grave{b},\grave{c},...=5,6,7$ can be
used for corresponding spacelike hyper surfaces on a base manifold and
typical cofiber.} In general frame form, such sets of geometric objects are
respectively denoted $\ ^{\shortmid }\mathbf{N}(\tau )=\ ^{\shortmid }%
\mathbf{N}(\tau ,\ ^{\shortmid }u)$ and$\ ^{\shortmid }\mathbf{g}(\tau )=\
^{\shortmid }\mathbf{g}(\tau ,\ ^{\shortmid }u).$ Let us write
correspondingly $\widetilde{\Xi }=(\widetilde{\Xi }_{t},\ \widetilde{\Xi }%
_{E})$ and $\ ^{\shortmid }\widetilde{\Xi }=(\widetilde{\Xi }_{t},\
^{\shortmid }\widetilde{\Xi }_{E})$ $\ $for nonholonomic distributions of
base and fiber hypersurfaces with conventional splitting 3+3 of signature
(+++;+++) on total phase space $T^{\ast }\mathbf{V.}$ \ On a typical cofiber
of such a phase space, we can consider a 3-d cofiber hypersurface $\
^{\shortmid }\widetilde{\Xi }_{E}$, for instance, of signature $(+++)$ with
a label $p_{8}=E$ for an energy type parameter.  Using N--adapted
(3+1)+(3+1) frame and coordinate transforms of metrics with additional
dependence on a flow parameter, we can parameterise the d-metric in the form%
\begin{eqnarray}
&&\ ^{\shortmid }\mathbf{g}=\ ^{\shortmid }\mathbf{g}_{\alpha ^{\prime
}\beta ^{\prime }}(\tau ,\ ^{\shortmid }u)d\ ^{\shortmid }\mathbf{e}^{\alpha
^{\prime }}\otimes d\ ^{\shortmid }\mathbf{e}^{\beta ^{\prime }}=q_{i}(\tau
,x^{k})dx^{i}\otimes dx^{i}+q_{3}(\tau ,x^{k},y^{3})\mathbf{e}^{3}\otimes
\mathbf{e}^{3}-[\breve{N}(\tau ,x^{k},y^{3})]^{2}\mathbf{e}^{4}\otimes
\mathbf{e}^{4}+  \notag \\
&&\ ^{\shortmid }q^{a_{2}}(\tau ,x^{k},y^{3},p_{b_{2}})\ ^{\shortmid }%
\mathbf{e}_{a_{2}}\otimes \ ^{\shortmid }\mathbf{e}_{a_{2}}+\ ^{\shortmid
}q^{7}(\tau ,x^{k},y^{3},p_{b_{2}},p_{b_{3}})\ ^{\shortmid }\mathbf{e}%
_{7}\otimes \ ^{\shortmid }\mathbf{e}_{7}-[\ ^{\shortmid }\check{N}(\tau
,x^{k},y^{3},p_{b_{2}},p_{b_{3}})]^{2}\ ^{\shortmid }\mathbf{e}_{8}\otimes \
^{\shortmid }\mathbf{e}_{8},  \label{31metric}
\end{eqnarray}%
where $\ ^{\shortmid }e^{\alpha _{s}}$ (for $a_{2}=5,6)$ are N-adapted
bases. This ansatz is written as an extension of a couple of 3--d metrics, $%
q_{ij}=diag(q_{\grave{\imath}})=(q_{i},q_{3})$ on a hypersurface $\widetilde{%
\Xi }_{t},$ and $\ ^{\shortmid }q^{\grave{a}\grave{b}}=diag(\ ^{\shortmid
}q^{\grave{a}})=(\ ^{\shortmid }q^{a_{2}},\ ^{\shortmid }q^{7})$ on a
hypersurface $\ ^{\shortmid }\widetilde{\Xi }_{E},$ \ if $\ q_{3}=g_{3},%
\breve{N}^{2}=-g_{4}$ and$\ ^{\shortmid }q^{7}=\ ^{\shortmid }g^{7},\
^{\shortmid }\check{N}^{2}=-\ ^{\shortmid }g^{8}.$ We consider $\breve{N}$
as the lapse function on the base and $\ ^{\shortmid }\check{N}$ as the
lapse function in the co-fiber. In this work, we elaborate on geometric
phase flow theories on a conventional temperature like parameter $\tau .$

The theory of geometric flows of relativistic Hamilton mechanical systems
was formulated \cite{vacaru19a} in explicit form using canonical data $(\
^{\shortmid }\widetilde{\mathbf{g}}(\tau ),\ ^{\shortmid }\widetilde{\mathbf{%
D}}(\tau )),$ in terms of geometric objects with "tilde" values. Considering
nonholonomic frame transforms and deformations of d-connections, and
redefining the normalizing functions, we can postulate such generalizations
of the G. Perelman functionals:
\begin{eqnarray}
\ ^{\shortmid }\mathcal{F} &=&\ ^{\shortmid }\int e^{-\ ^{\shortmid }f}\sqrt{%
|\ ^{\shortmid }\mathbf{g}_{\alpha \beta }|}d^{8}\ ^{\shortmid }u(\
_{s}^{\shortmid }R+|\ ^{\shortmid }\mathbf{D}\ ^{\shortmid }f|^{2})%
\mbox{
and }  \label{ffperelmctl} \\
\ ^{\shortmid }\mathcal{W} &=&\ ^{\shortmid }\int \ ^{\shortmid }\mu \sqrt{%
|\ ^{\shortmid }\mathbf{g}_{\alpha \beta }|}d^{8}\ ^{\shortmid }u[\tau (\ \
_{s}^{\shortmid }R+|\ \ _{h}^{\shortmid }\mathbf{D}\ ^{\shortmid }f|+|\ \
_{v}^{\shortmid }\mathbf{D}\ \ ^{\shortmid }f|)^{2}+\ \ ^{\shortmid }f-16].
\label{wfperelmctl}
\end{eqnarray}%
In these formulas, we use a brief notation for the integrals on phase space
variables and the normalizing function $\ \ ^{\shortmid }f(\tau ,\
^{\shortmid }u)$ is subjected to the conditions
\begin{equation*}
\ ^{\shortmid }\int \ ^{\shortmid }\mu \sqrt{|\ ^{\shortmid }\mathbf{g}%
_{\alpha \beta }|}d^{8}\ ^{\shortmid }u=\int_{t_{1}}^{t_{2}}\int_{\Xi
_{t}}\int_{E_{1}}^{E_{2}}\int_{\ ^{\shortmid }\Xi _{E}}\ ^{\shortmid }\mu
\sqrt{|\ ^{\shortmid }\mathbf{g}_{\alpha \beta }|}d^{8}\ ^{\shortmid }u=1
\end{equation*}%
for a classical integration measure$\ ^{\shortmid }\mu =\left( 4\pi \tau
\right) ^{-8}e^{-\ ^{\shortmid }f}$ and the Ricci scalar $\ _{s}^{\shortmid
}R$ is taken for the Ricci d-tensor $\ ^{\shortmid }\mathbf{R}_{\alpha \beta
}$ of a d-connection $\ ^{\shortmid }\mathbf{D}.$

Similar F- and W--functionals can be postulated for nonholonomic geometric
flows on $T^{\ast }\mathbf{V}$ using data $(\ ^{\shortmid }\mathbf{g}(\tau
),\ \ ^{\shortmid }\widehat{\mathbf{D}}(\tau )),$ or $(\ ^{\shortmid }%
\widetilde{\mathbf{g}}(\tau ),\ \ ^{\shortmid }\widetilde{\mathbf{D}}(\tau
)),$ and other type ones related via distorting relations, with
correspondingly redefined integration measures and normalizing functions,
and respective hypersurfaces. LC-configurations can be extracted for certain
conditions when $\ ^{\shortmid }\mathbf{D}_{\mid \ ^{\shortmid }\mathbf{T}%
=0}=\ ^{\shortmid }\nabla .$

\section{G. Perelman \& von Neumann entropies for geometric information flows}

\label{s3}Geometric flows of Riemannian metrics are characterized by a
statistical thermodynamic model which can be elaborated in a self-consistent
form using a W-functional of type (\ref{wfperelmctl}) defined for Riemannian
metrics which has properties of "minus entropy" \cite{perelman1}.
Introducing a respective thermodynamic generation function, all
thermodynamic values can be defined and computed by integrating
with corresponding measures defined by the metric structure and a
corresponding normalizing function. Similar constructions can be elaborated
for various relativistic, supersymmetric, commutative and noncommutative
generalizations if the geometric flow evolution is modelled for
corresponding nonholonomic fibered structures preserving causality and basic
postulates for self-consistent stochastic, kinetic and thermodynamics models
\cite{vacaru00ap,vacaru09,rajpoot17,ruchin13,gheorghiu16}, see also \cite%
{vacaru19,vacaru19a,vacaru19b,vacaru18,bubuianu18} and references therein.
Originally, such nonholonomic transforms of geometric objects and
deformations of the (non) linear connection structures were considered in
\cite{vjmp08,vrmp09} where the theory of geometric flows was generalized for
Finsler-Lagrange geometries. Then the approach was developed for flows of
Hamilton classical and quantum mechanical systems with certain applications
in information theory \cite{vacaru19b}. In this section, we redefine the
constructions changing the "mechanical" variables into general N-adapted
ones which is important for further developments in quantum information and
gravity theories.

\subsection{Analogous thermodynamic models for Hamiltonian GIFs}

For relativistic geometric flows of mechanical systems described by
Hamiltonians \cite{vacaru19a,vacaru19b}, the thermodynamic generating
function can be written in the form $\ \ ^{\shortmid }\widetilde{\mathcal{Z}}%
[\ ^{\shortmid }\widetilde{\mathbf{g}}(\tau )]=\ ^{\shortmid }\widetilde{%
\int }e^{-\ ^{\shortmid }\widetilde{f}}\sqrt{|\ ^{\shortmid }\widetilde{%
\mathbf{g}}_{\alpha \beta }|}d^{8}\ ^{\shortmid }u(-\ ^{\shortmid }%
\widetilde{f}+16)$, on $T^{\ast }\mathbf{V,}$ where the integral $\
^{\shortmid }\widetilde{\int }$ is considered for canonical mechanical
variables and the corresponding functional dependence is determined by $\
^{\shortmid }\widetilde{\mathbf{g}}(\tau ).$\footnote{%
Hereafter, we shall not write such dependencies in explicit form if that
will not result in ambiguities.} With respect to general frames (or with
necessary (3+1)+(3+1) decomposition and a d-metric of type (\ref{31metric}%
)), the integration measure can be re-defined in a form which allows us to consider
\begin{equation}
\ ^{\shortmid }\mathcal{Z}[\ ^{\shortmid }\mathbf{g}(\tau )]=\ ^{\shortmid
}\int e^{-\ ^{\shortmid }f}\sqrt{|\ ^{\shortmid }\mathbf{g}_{\alpha \beta }|}%
d^{8}\ ^{\shortmid }u(-\ ^{\shortmid }f+16),\mbox{ for }T^{\ast }\mathbf{V.}
\label{genarthermfunct}
\end{equation}%
A variational N-adapted calculus for $\ ^{\shortmid }\mathcal{Z}$ and
geometric data $(\ ^{\shortmid }\mathbf{N,}\ ^{\shortmid }\mathbf{g,}\
^{\shortmid }\mathbf{D})$ allows us to compute such relativistic
thermodynamic values:
\begin{eqnarray}
\ ^{\shortmid }\mathcal{E}\ &=&-\tau ^{2}\ ^{\shortmid }\int e^{-\
^{\shortmid }f}\sqrt{|q_{1}q_{2}q_{3}\breve{N}\ ^{\shortmid }q_{5}\
^{\shortmid }q_{6}\ ^{\shortmid }q_{7}\ ^{\shortmid }\check{N}|}\delta ^{8}\
^{\shortmid }u(\ \ _{s}^{\shortmid }R+|\ ^{\shortmid }\mathbf{D}\
^{\shortmid }f|^{2}\mathbf{\ }-\frac{8}{\tau }),  \label{8rdthvhs} \\
\ ^{\shortmid }\mathcal{S}\ &=&-\ ^{\shortmid }\int e^{-\ ^{\shortmid }f}%
\sqrt{|q_{1}q_{2}q_{3}\breve{N}\ ^{\shortmid }q_{5}\ ^{\shortmid }q_{6}\
^{\shortmid }q_{7}\ ^{\shortmid }\check{N}|}\delta ^{8}\ ^{\shortmid }u\left[
\tau \left( \ _{s}^{\shortmid }R+|\ ^{\shortmid }\mathbf{D}\ ^{\shortmid
}f|^{2}\right) +\ ^{\shortmid }f-16\right] ,  \notag \\
\ ^{\shortmid }\eta \ &=&-\ ^{\shortmid }\int e^{-\ ^{\shortmid }f}\sqrt{%
|q_{1}q_{2}q_{3}\breve{N}\ ^{\shortmid }q_{5}\ ^{\shortmid }q_{6}\
^{\shortmid }q_{7}\ ^{\shortmid }\check{N}|}\delta ^{8}\ ^{\shortmid }u[|\ \
^{\shortmid }\mathbf{R}_{\alpha \beta }+\ ^{\shortmid }\mathbf{D}_{\alpha }\
\ ^{\shortmid }\mathbf{D}_{\beta }\ ^{\shortmid }f-\frac{1}{2\tau }\
^{\shortmid }\mathbf{g}_{\alpha \beta }|^{2}].  \notag
\end{eqnarray}%
Such values can be written in Hamilton mechanical variables with tilde as in
(\ref{genarthermfunct}) or re-defining the normalizing functions for the
canonical d-connection $\ ^{\shortmid }\widehat{\mathbf{D}}\mathbf{,}$ see (%
\ref{prefercon}) and respective distorting relations.

Using the first two formulas in (\ref{8rdthvhs}) for two d-metrics $\ _{1}%
\mathbf{g}$ and $\mathbf{g}$, we can define  the respective free
energy and relative entropy,%
\begin{equation*}
\ ^{\shortmid }\mathcal{F}(\ _{1}\mathbf{g})=\mathcal{E}(\ _{1}\mathbf{g}%
)-\tau \ ^{\shortmid }\mathcal{S}(\ _{1}\mathbf{g})\mbox{ and }\ ^{\shortmid
}\mathcal{S}(\ _{1}\mathbf{g}\shortparallel \mathbf{g})=\beta \lbrack \
^{\shortmid }\mathcal{F}(\ _{1}\mathbf{g})-\ ^{\shortmid }\mathcal{F}(%
\mathbf{g})],\mbox{ where }
\end{equation*}%
\begin{eqnarray*}
\mathcal{E}(\ _{1}\mathbf{g}) &=&-\tau ^{2}\int e^{-f}\sqrt{|q_{1}q_{2}q_{3}%
\breve{N}q_{5}q_{6}q_{7}\check{N}|}\delta ^{8}u[\ _{s}R(\ _{1}\mathbf{g})+|%
\mathbf{D}(\ _{1}\mathbf{g})f(\tau ,u)|^{2}\mathbf{\ }-\frac{8}{\tau }], \\
\mathcal{S}(\ _{1}\mathbf{g}) &=&-\int e^{-f}\sqrt{|q_{1}q_{2}q_{3}\breve{N}%
q_{5}q_{6}q_{7}\check{N}|}\delta ^{8}u\left[ \tau \left( \ _{s}R(\ _{1}%
\mathbf{g})+|\mathbf{D}(\ _{1}\mathbf{g})f(\tau ,u)|^{2}\right) +f(\tau
,u)-16\right]
\end{eqnarray*}%
are computed using the phase spacetime measures, the Ricci scalar and
canonical d--connection are determined respectively by $\ \mathbf{g}$ and $\
_{1}\mathbf{g,}$

In this work, we study the geometric flow evolution of thermodynamics
systems that preserves the thermal equilibrium at temperature $\beta $ for
maps $\ _{1}\mathbf{g}\rightarrow \ _{2}\mathbf{g.}$ A realistic physical
interpretation for such systems exists if
\begin{equation}
\mathcal{S}(\ _{1}\mathbf{g}\shortparallel \mathbf{g})\geq \mathcal{S}(\ _{2}%
\mathbf{g}\shortparallel \mathbf{g}),\mbox{ i.e. }\mathcal{F}(\ _{1}\mathbf{g%
})\geq \mathcal{F}(\ _{2}\mathbf{g}).  \label{secondthlaw}
\end{equation}
These aspects connect general frame and mechanical variables flow models to
the second low of thermodynamics. Values of type (\ref{8rdthvhs}) are in
relativistic thermodynamic relation if the second thermodynamic law (\ref%
{secondthlaw}) is satisfied. Such conditions impose additional constraints
on the class of normalizing and generating functions.

\subsection{Density matrix and entropies for quantum information flows}

\label{ssdmgif}In this subsection, we develop the density matrix formalism
for applications in the theory of classical and quantum geometric
information flows (respectively, GIFs and QGIFs), see sections 4 and 5 in
\cite{vacaru19b} for a formulation in Hamilton mechanical variables.
Nonholonomic deformations of G. Perelman entropy like functionals will be
used for relativistic formulations of the von Neumann entropy and QGIFs in
arbitrary frames.

\subsubsection{Statistical density matrix for relativistic classical GIFs}

The thermodynamic generating function $\ ^{\shortmid }\mathcal{Z}[\
^{\shortmid }\mathbf{g}(\tau )]$ (\ref{genarthermfunct}) with free energy $\
^{\shortmid }\mathcal{E}\ $can be used for defining the state density
\begin{equation}
\ ^{\shortmid }\sigma (\beta ,\ ^{\shortmid }\mathcal{E}\ ,\ ^{\shortmid }%
\mathbf{g})=\ ^{\shortmid }\mathcal{Z}^{-1}e^{-\beta \ ^{\shortmid }\mathcal{%
E}\ },  \label{statedens}
\end{equation}%
with $\beta =1/T,$ $\tau =T.$ This value is the a classical analog of the
density matrix in QM. We shall use it for elaborating models of QGIFs.

We can consider that a density state $\ ^{\shortmid }\sigma \lbrack \
^{\shortmid }\mathbf{g}]$ is associated to $\ ^{\shortmid }\mathbf{g}%
_{\alpha \beta },$ when but the geometric evolution may involve another
density $\ ^{\shortmid }\rho \lbrack \ _{1}^{\shortmid }\widetilde{\mathbf{g}%
}],$ where the left label 1 is used for distinguishing two d-metrics $\
^{\shortmid }\mathbf{g}$ and $\ \ _{1}^{\shortmid }\mathbf{g}.$ In result,
the concept of \textit{relative entropy} between any state density$\
^{\shortmid }\rho (\beta ,\ _{1}^{\shortmid }\mathcal{E}\ ,\ _{1}^{\shortmid
}\mathbf{g})$ and $\ ^{\shortmid }\widetilde{\sigma }(\beta ,\ ^{\shortmid }%
\mathcal{E}\ ,\ ^{\shortmid }\mathbf{g})$ can be introduced. It can be
computed for a prescribed measure $\omega (E)$ on a cotangent Lorentz bundle
with $E$ considered as a thermodynamical energy parameter associated to $\
^{\shortmid }\mathcal{E}.$

The \textit{conditional entropy} for GIFs is introduced%
\begin{equation}
\ ^{\shortmid }\mathcal{S}(\ ^{\shortmid }\rho \shortparallel \ ^{\shortmid
}\sigma )=\beta \lbrack \ ^{\shortmid }\mathcal{F}(\ ^{\shortmid }\rho )-\
^{\shortmid }\mathcal{F}(\ ^{\shortmid }\sigma )],  \label{condhamentr}
\end{equation}%
where the \textit{free energy }corresponding to $\ ^{\shortmid }\rho $ is
defined by formula
\begin{equation*}
\ ^{\shortmid }\mathcal{F}(\ ^{\shortmid }\rho ):=\ ^{\shortmid }\mathcal{E}%
(\ ^{\shortmid }\rho )-T\ ^{\shortmid }\mathcal{S}(\ ^{\shortmid }\rho )
\end{equation*}
with the \textit{average energy} $\ ^{\shortmid }\mathcal{E}(\ ^{\shortmid
}\rho )=\int \ ^{\shortmid }\rho Ed\omega (E).$ The thermodynamic entropy in
(\ref{condhamentr}) is computed following formula
\begin{equation*}
\ ^{\shortmid }\mathcal{S}(\ ^{\shortmid }\rho ):=\beta \ ^{\shortmid }%
\mathcal{E}(\ ^{\shortmid }\rho )+\log \ ^{\shortmid }\mathcal{Z}(\
^{\shortmid }\rho ).
\end{equation*}
The condition $\ ^{\shortmid }\mathcal{S}(\ ^{\shortmid }\sigma
\shortparallel \ ^{\shortmid }\sigma )=0$ is satisfied if $\log \
^{\shortmid }\mathcal{Z}$ is independent on $\ ^{\shortmid }\rho .$

\subsubsection{Entanglement and density matrix for QGIFs}

Using canonical mechanical variables $(\widetilde{H},\ ^{\shortmid }%
\widetilde{g}^{ab}),$ we can study special QM systems described by pure
states. In a more general context, QM involves probabilities considered not for a
quantum state but for densities matrices. In this subsection, we elaborate on
how GIFs of classical mechanical systems can be generalized to QGIFs using
basic concepts of quantum mechanics, QM, and information theory. We shall
elaborate on quantum models of GIFs described in terms of density matrices
defined as quantum analogs of state densities of type $\ ^{\shortmid }\sigma
$ (\ref{statedens}).

For any point $\ ^{\shortmid }u\in T^{\ast }\mathbf{V}$ of a typical
relativistic phase space used for modeling a classical GIF system $\mathcal{A%
}=\left[ \ ^{\shortmid }\mathcal{E},\ ^{\shortmid }\mathcal{S},\ ^{\shortmid
}\eta \right] $ (\ref{8rdthvhs}), we associate a typical Hilbert space $%
\mathcal{H}_{\mathcal{A}},$ which is denoted $\widetilde{\mathcal{H}}_{%
\mathcal{A}}$ for canonical Hamilton mechanical variables. A state vector $%
\widetilde{\psi }_{\mathcal{A}}\in \widetilde{\mathcal{H}}_{\mathcal{A}}$
can be defined as an infinite dimensional complex vector function. For
applications in quantum information theory, there are considered
approximations with finite dimensions. Such a $\widetilde{\psi }_{\mathcal{A}%
}$ is a solution of the Schr\"{o}dinger equation with a Hamiltonian $%
\widehat{H}$ \ constructed as a well-defined quantum version of a canonical
Hamiltonian $\widetilde{H}.$ In a a more general context, we can work with
general covariant variables (or certain versions with (3+1)+(3+1)
splitting), when "non-tilde" d-metrics $\ ^{\shortmid }\mathbf{g}$ (see (\ref%
{31metric})) are used for definition of certain quantum measures.
Considering unitary transforms of type $\widetilde{\psi }_{\mathcal{A}%
}\rightarrow U$ $\psi _{\mathcal{A}},$ we can describe the system $A$ by an
abstract Hilbert space $\mathcal{H}_{\mathcal{A}},$ or to associate a
complex vector space of dimension $\underline{N}$ with Hermitian product,
see details in \cite{preskill,witten18}.

The complex geometric arena for QGIFs models consists from complex bundles $%
\mathcal{H}_{\mathcal{A}}(T^{\ast }\mathbf{V})=\cup _{\ ^{\shortmid }u}%
\mathcal{H}_{\mathcal{A}}^{\ ^{\shortmid }u}$ assoctiated to $T^{\ast }%
\mathbf{V}$ and constructed as unities of Hilbert spaces $\mathcal{H}_{%
\mathcal{A}}^{\ ^{\shortmid }u}$ for $\ ^{\shortmid }u\in T^{\ast }\mathbf{V,%
}$ or a points of a subspace of such a phase space. We consider that there
are nonholonomic variables when $\psi _{\mathcal{A}}\rightarrow \widetilde{%
\psi }_{\mathcal{A}}(\ ^{\shortmid }u)$ and the integration measure is
determined by a$\ ^{\shortmid }\widetilde{\mathbf{g}},$ or its frame
transforms to a $\ ^{\shortmid }\mathbf{g.}$ It is assumed that such
constructions are possible at least for perturbanions nearly a flat "double"
Minkowski metric (\ref{lqed}) nearly a point $\ ^{\shortmid }u.$ This way a
perturbative QGIF model with quasi-classical limits can be always
elaborated. GIFs describe flow evolution of mechanical systems in causal relativistic classical
forms.

The combined Hilbert space is defined as a tensor product, $\mathcal{H}_{%
\mathcal{A}}\otimes \mathcal{H}_{\mathcal{B}}$, with an associate Hilbert
space $\mathcal{H}_{\mathcal{A}}$ considered for a complementary system $%
\mathcal{A}.$ Here we note that symbols $\mathcal{A},\mathcal{B},\mathcal{C}$
etc. are used as labels for certain systems under geometric evolutions
described by respective thermodynamical modles \ of type (\ref{8rdthvhs}).
The state vectors for a combined QGIF system are written $\psi _{\mathcal{AB}%
}=\psi _{\mathcal{A}}\otimes \psi _{\mathcal{B}}\in \mathcal{H}_{\mathcal{AB}%
}=\mathcal{H}_{\mathcal{A}}\otimes \mathcal{H}_{\mathcal{B}}$ for $\psi _{%
\mathcal{A}}=1_{\mathcal{A}}$ taken as the unity state vector. Quantum
systems subjected only to quantum evolution and not to geometric flows are
denoted $A,B,C,...$

\paragraph{Entangled states:}

In QM and QGIF theories, a pure state $\psi _{\mathcal{AB}}\in \mathcal{H}_{%
\mathcal{AB}}$ may be not only a tensor product vector but also \textit{%
entangled} and represented by a matrix of dimension $\underline{N}\times
\underline{M}$ if $\dim \mathcal{H}_{A}=\underline{N}$ and $\dim \mathcal{H}%
_{B}=\underline{M}$. We underline such symbols in order to avoid ambiguities
with the N-connection symbol $N$. A Schmidt decomposition can be considered
for any pure state,
\begin{equation}
\psi _{\mathcal{AB}}=\sum_{\underline{i}}\sqrt{p_{\underline{i}}}\psi _{%
\mathcal{A}}^{\underline{i}}\otimes \psi _{\mathcal{B}}^{\underline{i}},
\label{schm1}
\end{equation}
for any index $\underline{i}=1,2,....$(up to a finite value). The state
vectors $\psi _{\mathcal{A}}^{\underline{i}}$ can be taken to be
orthonormal, $<\psi _{\mathcal{A}}^{\underline{i}},\psi _{\mathcal{A}}^{%
\underline{j}}>=<\psi _{\mathcal{B}}^{\underline{i}},\psi _{\mathcal{B}}^{%
\underline{j}}>=\delta ^{\underline{i}\underline{j}}$, where $\delta ^{%
\underline{i}\underline{j}}$ is the Kronecker symbol. If $p_{\underline{i}%
}>0 $ and $\sum_{\underline{i}}p_{\underline{i}}=1,$ we can treat $p_{%
\underline{i}}$ as probabilities. In general, such $\psi _{\mathcal{A}}^{%
\underline{i}}$ and/or $\psi _{\mathcal{B}}^{\underline{i}}$ do not define
bases of $\mathcal{H}_{A}$ and/or $\mathcal{H}_{B}$ because we can take some
vectors when, in principle, it is not enough for such bases. We can consider
aht such values split the GIFs into certain probable evolution scenarios.

\paragraph{The quantum density matrix}

for a QGIF-associated system $\mathcal{A}$ is defined
\begin{equation}
\rho _{\mathcal{A}}:=\sum_{\underline{a}}p_{\underline{a}}|\psi _{\mathcal{A}%
}^{\underline{a}}><\otimes \psi _{\mathcal{A}}^{\underline{a}}|
\label{schm2}
\end{equation}%
as a Hermitian and positive semi-definite operator with trace $Tr_{\mathcal{H%
}_{\mathcal{A}}}\rho _{\mathcal{A}}=1.$ Using such a $\rho _{\mathcal{A}},$
we can compute the \textit{expectation} value of any operator $\mathcal{O}_{%
\mathcal{A}}$ characterizing additionally such a system,%
\begin{eqnarray}
<\mathcal{O}>_{\mathcal{AB}} &=&<\psi _{\mathcal{AB}}|\mathcal{O}_{\mathcal{A%
}}\otimes 1_{\mathcal{B}}|\psi _{\mathcal{AB}}>=\sum_{\underline{i}}p_{%
\underline{i}}<\psi _{\mathcal{A}}^{\underline{i}}|\mathcal{O}_{\mathcal{A}%
}|\psi _{\mathcal{A}}^{\underline{i}}><\psi _{\mathcal{B}}^{\underline{i}%
}|1_{\mathcal{B}}|\psi _{\mathcal{B}}^{\underline{i}}>=  \notag \\
<\mathcal{O}>_{\mathcal{A}} &=&\sum_{\underline{i}}p_{\underline{i}}<\psi _{%
\mathcal{A}}^{\underline{i}}|\mathcal{O}_{\mathcal{A}}|\psi _{\mathcal{A}}^{%
\underline{i}}>=Tr_{\mathcal{H}_{\mathcal{A}}}\rho _{\mathcal{A}}\mathcal{O}%
_{\mathcal{A}}.  \label{expectvalues}
\end{eqnarray}%
Such values encode both quantum information and geometric flow evolution of
bipartite systems of type $\mathcal{A},\mathcal{B},$ and $\mathcal{AB}$ with
both quantum and geometric entanglement defined by density matrices.

\paragraph{Joint probabilities for bipartite quantum systems and measurements:}

Bipartite QGIFs systems are described in general form by quantum density
matrices of type $\rho _{\mathcal{AB}}$ or (in canonical mechanical
variables) $\rho _{\widetilde{\mathcal{A}}\widetilde{\mathcal{B}}}.$ In the
classical probability theory, we describe a bipartite system $XY$ by a
\textit{joint probability} distribution $P_{X,Y}(x_{\underline{i}},y_{%
\underline{j}}),$ where $P_{X}(x_{\underline{i}}):=\sum_{\underline{j}%
}P_{X,Y}(x_{\underline{i}},y_{\underline{j}}),$ see details in \cite%
{preskill,witten18} and, for GIFs, \cite{vacaru19b}.

Considering $\mathcal{AB}$ as a bipartite quantum system with Hilbert space $%
\mathcal{H}_{\mathcal{A}}\otimes \mathcal{H}_{\mathcal{B}},$ we can define
and parameterize a QGIF density matrix $\rho _{\mathcal{AB}}$ in standard QM
form:
\begin{equation*}
\rho _{\mathcal{AB}}=\sum_{\underline{a},\underline{a}^{\prime },\underline{b%
},\underline{b}^{\prime }}\rho _{\underline{a}\underline{a}^{\prime }%
\underline{b}\underline{b}^{\prime }}|\underline{a}>_{\mathcal{A}}\otimes |%
\underline{b}>_{\mathcal{B}}\ _{\mathcal{A}}<\underline{a}^{\prime }|\otimes
\ _{\mathcal{B}}<\underline{b}^{\prime }|.
\end{equation*}%
In this formula, $|\underline{a}>_{A},$ $\underline{a}=1,2,...,$\underline{$%
n $} is an orthonormal basis of $\mathcal{H}_{\mathcal{A}}$ and $|\underline{%
b}>_{\mathcal{B}},$ $\underline{b}=1,2,...,$\underline{$m$} as an
orthonormal basis of $\mathcal{H}_{\mathcal{B}}.$

A \textit{measurement} of the system $\mathcal{A}$ is characterized by a
\textit{reduced density matrix} obtained by respective contracting of
indices,
\begin{equation*}
\rho _{\mathcal{A}}=Tr_{\mathcal{H}_{\mathcal{B}}}\rho _{\mathcal{AB}}=\sum_{%
\underline{a},\underline{a}^{\prime },\underline{b},\underline{b}}\rho _{%
\underline{a}\underline{a}^{\prime }\underline{b}\underline{b}}|\underline{a}%
>_{\mathcal{A}}\ _{\mathcal{A}}<\underline{a}^{\prime }|,\mbox{ for }|%
\underline{b}>_{\mathcal{B}}\ _{\mathcal{B}}<\underline{b}|=1.
\end{equation*}%
In a similar form, we can define and compute $\rho _{\mathcal{B}}=Tr_{%
\mathcal{H}_{\mathcal{A}}}\rho _{\mathcal{AB}}.$ For cotangent bundle
constructions, we can distinguish the geometric and physical objects putting
left labels "$\ ^{\shortmid }$", $\ ^{\shortmid }\rho _{\mathcal{B}}=Tr_{%
\mathcal{H}_{\mathcal{A}}}\ ^{\shortmid }\rho _{\mathcal{AB}}.$ Using such
formulas, we can elaborate on QGIFs models and quantum information theory
formulated in conventional mechanical variables or in a general covariant
form.

\subsubsection{Quantum density matrix for QGIFs}

The quantum density matrix $\ ^{\shortmid }\sigma _{\mathcal{AB}}$ for a
state density $\ ^{\shortmid }\sigma $ (\ref{statedens}) can be defined and
computed using formulas (\ref{expectvalues}),
\begin{eqnarray}
\ ^{\shortmid }\sigma _{\mathcal{AB}} &=&<\ ^{\shortmid }\sigma >_{\mathcal{%
AB}}=<\psi _{\mathcal{AB}}|\ ^{\shortmid }\sigma \otimes 1_{\mathcal{B}%
}|\psi _{\mathcal{AB}}>=\sum_{\underline{i}}p_{\underline{i}}<\psi _{%
\mathcal{A}}^{\underline{i}}|\ ^{\shortmid }\sigma |\psi _{\mathcal{A}}^{%
\underline{i}}><\psi _{\mathcal{B}}^{\underline{i}}|1_{\mathcal{B}}|\psi _{%
\mathcal{B}}^{\underline{i}}>=  \notag \\
\ ^{\shortmid }\sigma _{\mathcal{A}} &=&<\ ^{\shortmid }\sigma >_{\mathcal{A}%
}=\sum_{\underline{i}}p_{\underline{i}}<\psi _{\mathcal{A}}^{\underline{i}%
}|\ ^{\shortmid }\sigma |\psi _{\mathcal{A}}^{\underline{i}}>=Tr_{\mathcal{H}%
_{\mathcal{A}}}\ ^{\shortmid }\rho _{\mathcal{A}}\ ^{\shortmid }\sigma ,
\label{aux01}
\end{eqnarray}%
where the density matrix $\ ^{\shortmid }\rho _{\mathcal{A}}$ is taken for
computing the QGIF density matrix $\ ^{\shortmid }\sigma _{\mathcal{A}}.$
This matrix is determined by a state density of the thermodynamical model
for GIFs of a classical system $\ ^{\shortmid }\sigma $ which can be
parameterized in nonholonomic variables of a mechanical Hamiltonian system $%
\ ^{\shortmid }\widetilde{\sigma }.$

For quantum systems, we can work with quantum density matrices $\
^{\shortmid }\sigma _{\mathcal{AB}}$ and $\ ^{\shortmid }\sigma _{\mathcal{A}%
}$ and respective partial traces $\ ^{\shortmid }\sigma _{\mathcal{A}}=Tr_{%
\mathcal{H}_{\mathcal{B}}}\ ^{\shortmid }\sigma _{\mathcal{AB}}$ and $\
^{\shortmid }\sigma _{\mathcal{B}}=Tr_{\mathcal{H}_{\mathcal{A}}}\
^{\shortmid }\sigma _{\mathcal{AB}}.$ Such formulas can be written in
coefficient forms
\begin{equation*}
\ ^{\shortmid }\sigma _{\mathcal{AB}}=\sum_{\underline{a},\underline{a}%
^{\prime },\underline{b},\underline{b}^{\prime }}\ ^{\shortmid }\sigma _{%
\underline{a}\underline{a}^{\prime }\underline{b}\underline{b}^{\prime }}|%
\underline{a}>_{\mathcal{A}}\otimes |\underline{b}>_{\mathcal{B}}\ _{%
\mathcal{A}}<\underline{a}^{\prime }|\otimes \ _{\mathcal{B}}<\underline{b}%
^{\prime }|\mbox{ and }\ ^{\shortmid }\sigma _{\mathcal{A}}=\sum_{\underline{%
a},\underline{a}^{\prime },\underline{b},\underline{b}}\ ^{\shortmid }\sigma
_{\underline{a}\underline{a}^{\prime }\underline{b}\underline{b}}|\underline{%
a}>_{\mathcal{A}}\ _{\mathcal{A}}<\underline{a}^{\prime }|.
\end{equation*}%
Using a density matrix encoding the data for QGIFs of Hamilton mechanical
system described in general covariant variables, we can compute respective
thermodynamical values.

\subsubsection{The von Neumann entropy and QGIFs}

QGIFs can be described in standard QM form for the von Neumann entropy
determined by $\ ^{\shortmid }\sigma _{\mathcal{A}}$ (\ref{aux01}) \ as a
probability distribution,%
\begin{equation}
\ _{q}^{\shortmid }\mathcal{S}(\ ^{\shortmid }\sigma _{\mathcal{A}}):=Tr\
^{\shortmid }\sigma _{\mathcal{A}}\log \ ^{\shortmid }\sigma _{\mathcal{A}}.
\label{neumgfentr}
\end{equation}%
Hereafter we shall write the trace in a simplified form without a label for
the corresponding Hilbert space if that will not result in ambiguities. We
use also a left label $q$ to state the quantum character of such values. It
should be also emphasized that such an entropy is a quantum analog of a$\
^{\shortmid }\widetilde{\mathcal{S}}$ used in the thermodynamic model for
geometric flow evolution of Hamilton mechanical systems. Tilde can be
omitted for general frame transforms when $\ ^{\shortmid }\mathcal{S}$
encode a different frame structure. Such a QGIF entropy satisfies two
conditions: $\ _{q}^{\shortmid }\mathcal{S}(\ ^{\shortmid }\sigma _{\mathcal{%
A}})\geq 0$ and it is manifestly invariant under a unitary transformation $\
^{\shortmid }\sigma _{\mathcal{A}}\rightarrow U\ ^{\shortmid }\sigma _{%
\mathcal{A}}U^{-1}.$

The von Neuman entropy for QGIFs, $\ _{q}^{\shortmid }\mathcal{S}(\
^{\shortmid }\sigma _{\mathcal{A}}),$ has a purifying property which does
not have a classical analog. Considering a bipartite system $\psi _{\mathcal{%
AB}}=\sum_{\underline{i}}\sqrt{p_{\underline{i}}}\psi _{\mathcal{A}}^{%
\underline{i}}\otimes \psi _{\mathcal{B}}^{\underline{i}}$ and $\rho _{%
\mathcal{A}}:=\sum_{\underline{i}}p_{\underline{i}}|\psi _{\mathcal{A}}^{%
\underline{i}}>\otimes <\psi _{\mathcal{A}}^{\underline{i}}|,$ we compute
\begin{equation}
\ ^{\shortmid }\sigma _{\mathcal{A}}:=\sum_{\underline{a},\underline{a}%
^{\prime },\underline{b},\underline{b}}\ \sum_{\underline{k}}^{\shortmid
}\sigma _{\underline{a}\underline{a}^{\prime }\underline{b}\underline{b}}p_{%
\underline{k}}\ _{\mathcal{A}}<\underline{a}^{\prime }||\psi _{\mathcal{A}}^{%
\underline{k}}><\otimes \psi _{\mathcal{A}}^{\underline{k}}||\underline{a}>_{%
\mathcal{A}},\ ^{\shortmid }\sigma _{\mathcal{B}}:=\sum_{\underline{a},%
\underline{a}^{\prime },\underline{b},\underline{b}}\ \sum_{\underline{k}%
}^{\shortmid }\sigma _{\underline{a}\underline{a}^{\prime }\underline{b}%
\underline{b}}p_{\underline{k}}\ _{\mathcal{B}}<\underline{a}^{\prime
}||\psi _{\mathcal{B}}^{\underline{k}}><\otimes \psi _{\mathcal{B}}^{%
\underline{k}}||\underline{b}>_{\mathcal{B}}.  \label{aux03}
\end{equation}%
In these formulas, we have the same probabilities $p_{\underline{k}}\ $ for
two formulas with different matrices and bases. This proves that $\
_{q}^{\shortmid }\mathcal{S}(\ ^{\shortmid }\sigma _{\mathcal{A}})=\
_{q}^{\shortmid }\mathcal{S}(\ ^{\shortmid }\sigma _{\mathcal{B}})$ when a
system $\mathcal{A}$ and a purifying system $\mathcal{B}$ have the same von
Neumann entropy.

\subsubsection{Quantum generalizations of the W- and thermodynamic entropy}

QGIFs can be characterized not only by a von Neumann entropy of type (\ref%
{neumgfentr}) but also by quantum analogs of entropy values used for
classical geometric flows. We can consider both an associated
thermodynamics entropy and a W-entropy in classical variants and then quantize
such systems using a respective Hamiltonian which allows a self-consistent QM
formulation. Such values can be introduced and computed in explicit form
using respective formulas (\ref{aux01}), (\ref{aux03}) for classical
conditional (\ref{condhamentr}) and mutual entropy considered for GIFs and
in information theory \cite{preskill,witten18,vacaru19b}. We define
respectively%
\begin{eqnarray*}
\ _{q}^{\shortmid }\mathcal{W}_{\mathcal{AB}} &=&Tr_{\mathcal{H}_{\mathcal{AB%
}}}[(\ ^{\shortmid }\sigma _{\mathcal{AB}})(\ _{\mathcal{AB}}^{\shortmid }%
\mathcal{W})]\mbox{ and }\ _{q}^{\shortmid }\mathcal{W}_{\mathcal{A}}=Tr_{%
\mathcal{H}_{\mathcal{A}}}[(\ ^{\shortmid }\sigma _{\mathcal{A}})(\ _{%
\mathcal{A}}^{\shortmid }\mathcal{W})],\ _{q}^{\shortmid }\mathcal{W}_{%
\mathcal{B}}=Tr_{\mathcal{H}_{\mathcal{B}}}[(\ ^{\shortmid }\sigma _{%
\mathcal{B}})(\ _{\mathcal{B}}^{\shortmid }\mathcal{W})]; \\
\ _{q}^{\shortmid }\mathcal{S}_{\mathcal{AB}} &=&Tr_{\mathcal{H}_{\mathcal{AB%
}}}[(\ ^{\shortmid }\sigma _{\mathcal{AB}})(\ _{\mathcal{AB}}^{\shortmid }%
\mathcal{S})]\mbox{ and }\ _{q}^{\shortmid }\mathcal{S}_{\mathcal{A}}=Tr_{%
\mathcal{H}_{\mathcal{A}}}[(\ ^{\shortmid }\sigma _{\mathcal{A}})(\ _{%
\mathcal{A}}^{\shortmid }\mathcal{S})],\ _{q}^{\shortmid }\mathcal{S}_{%
\mathcal{B}}=Tr_{\mathcal{H}_{\mathcal{B}}}[(\ ^{\shortmid }\sigma _{%
\mathcal{B}})(\ _{\mathcal{B}}^{\shortmid }\mathcal{S})].
\end{eqnarray*}%
Such values describe corresponding entropic properties of quantum systems
with rich geometric structure under geometric flow evolution.

The quantum probabilistic characteristics are described by the von Neumann
entropy $\ _{q}^{\shortmid }\mathcal{S}(\ ^{\shortmid }\sigma _{\mathcal{A}%
}) $ (\ref{neumgfentr}) and corresponding generalizations for $\mathcal{AB}$
and $\mathcal{B}$ systems
\begin{equation*}
\ _{q}^{\shortmid }\mathcal{S}(\ ^{\shortmid }\sigma _{\mathcal{AB}}):=Tr\
^{\shortmid }\sigma _{\mathcal{AB}}\log \ ^{\shortmid }\sigma _{\mathcal{AB}}%
\mbox{ and }\ _{q}^{\shortmid }\mathcal{S}(\ ^{\shortmid }\sigma _{\mathcal{A%
}}):=Tr\ ^{\shortmid }\sigma _{\mathcal{A}}\log \ ^{\shortmid }\sigma _{%
\mathcal{A}},\ _{q}^{\shortmid }\mathcal{S}(\ ^{\shortmid }\sigma _{\mathcal{%
B}}):=Tr\ ^{\shortmid }\sigma _{\mathcal{B}}\log \ ^{\shortmid }\sigma _{%
\mathcal{B}}.
\end{equation*}%
Such values also encode thermodynamic, geometric flow and probabilistic
properties of QGIFs and can be used for elaborating a standard approach to
quantum information theory for systems with geometric mechanical Hamilton
flows and their covariant frame transforms.

\section{Entanglement and QGIFs of quantum mechanical systems}

\label{s4}Originally, the notion of bipartite entanglement was introduced for pure states and density matrix generalizations in description of finite-dimensional QM systems, see review of results in
\cite{preskill,witten18,aolita14,nishioka18}. In this section, we analyze how the concept of entanglement can be generalized for QGIFs when, for instance, there are considered two relativistic mechanical systems under geometric flow evolution. Such systems and their thermodynamic and QM analogs are characterized by a set of entropies like G. Perelman's W-entropy and geometric thermodynamic entropy and the nontrivial entanglement entropy in the von Neumann sense. Each of such entropic values characterise classical
and quantum correlations determined by geometric flow evolution and quantifies the amount of quantum entanglement. A set of inequalities involving Pereleman and entanglement entropies play a crucial role in
definition and description of such systems. We provide such formulas without rigorous proofs following two reasons: The W-entropy $\ ^{\shortmid }\mathcal{W}$ (\ref{wfperelmctl}), thermodynamic entropy
$\ ^{\shortmid }\mathcal{S}$ (\ref{8rdthvhs}) and related von Neumann
$\ _{q}^{\shortmid }\mathcal{S}$  (\ref{neumgfentr}) realizations are well-defined classical and
quantum entropic type values. For physicists, such formulas have a natural and intuitive motivation and interpretation in terms of thermodynamical generation functions and density matrices for GIFs. Rigorous mathematical proofs on hundreds of papers use methods of geometric analysis
\cite{perelman1,monogrrf1,monogrrf2,monogrrf3}. On main ideas and key steps for checking such results and selecting causal and realistic physical scenarios, we discuss in footnote 10 of our partner work \cite{vacaru19b}.

\subsection{Geometric flows with entanglement}

The goal of this subsection is to study how the concept of quantum entanglement can be developed for QGIF systems characterized by an associated statistical thermodynamic model with respective generating
function which transforms into a respective density matrix in a related quantum theory.

\subsubsection{Bipartite entanglement for QGIFs}

For any (relativistic) mechanic model, continuous or a lattice model of quantum field theory, thermo-field theory, QGIF model etc., we can associate a QM mechanical model with a pure ground state $|\Psi >$ for a total Hilbert space $\ _{t}\mathcal{H}$  when the density matrix is%
\begin{equation}
\ _{t}^{\shortmid }\rho =|\Psi ><\Psi |  \label{pureground}
\end{equation}%
can be normalized following the conditions $<\Psi |$ $\Psi >=1$ so that the
total trace $\ _{t}tr(\ _{t}^{\shortmid }\rho )=1.$ Such a conventional
total quantum system is divided into a two subsystems $\mathcal{A}$ and $%
\mathcal{B}.$ In this section, we consider that $\mathcal{A}=\left[ \
^{\shortmid }\mathcal{E},\ ^{\shortmid }\mathcal{S},\ ^{\shortmid }\eta %
\right] $ (\ref{8rdthvhs}) is a typical GIF system (in mechanical, or
general covariant variables) for with a QGIF model is elaborated. A similar
model (in principle, for a different associated relativistic Hamiltonian and
d-metric $\ _{1}\mathbf{g}$) is considered for $\mathcal{B}=\left[ \
_{1}^{\shortmid }\mathcal{E}, \ _{1}^{\shortmid }\mathcal{S},\
_{1}^{\shortmid }\eta \right] .$ Such subsystems $\mathcal{A}$ and $\mathcal{%
B}=\overline{\mathcal{A}}$ are complimentary to each other if in a $2n$%
-dimensional cotangent bundle space there is a common boundary $\partial
\mathcal{A}=\partial \mathcal{B}$ of codimension 2, where the non-singular
geometric flow evolution $\mathcal{A}$ transforms into a necessary analytic
class of flows on $\overline{\mathcal{A}}.$ In principle, we can consider
two completely different and classically separated GIF systems $\mathcal{A}$
and $\mathcal{B}$ which are correlated as quantum systems. We can consider
that for bipartite QGIFs $\ _{t}\mathcal{H=H}_{\mathcal{AB}}=\mathcal{H}_{%
\mathcal{A}}\otimes \mathcal{H}_{\mathcal{B}}$ as we considered in
subsection \ref{ssdmgif}. Such an approximation is less suitable, for
instance, if there are considered theories with gauge symmetries, see
discussion and references in footnote 3 of \cite{nishioka18} (we omit such
constructions in this work).

The measure of entanglement of a QGIF subsystem $\mathcal{A}$ is just the
von Neumann entropy $\ _{q}^{\shortmid }\mathcal{S}$ (\ref{neumgfentr}) but
defined for the reduced density matrix $\ ^{\shortmid }\rho _{\mathcal{A}%
}=Tr_{\mathcal{H}_{\mathcal{B}}}(\ _{t}^{\shortmid }\rho ),$ when the
\textit{entanglement entropy} of $\mathcal{A}$ is

\begin{equation}
\ _{q}^{\shortmid }\mathcal{S}(\ ^{\shortmid }\rho _{\mathcal{A}}):= Tr (\
^{\shortmid }\rho _{\mathcal{A}}\ \log \ ^{\shortmid }\rho _{\mathcal{A}}).
\label{entangentr}
\end{equation}%
Such a $\ ^{\shortmid }\rho _{\mathcal{A}}$ is associated to a state density
$\ ^{\shortmid }\rho (\beta ,\ ^{\shortmid }\mathcal{E}\ ,\ ^{\shortmid }%
\mathbf{g})$ of type (\ref{statedens}). We note that the total entropy $\
_{t}^{\shortmid }\mathcal{S}=0$ for a pure grand state (\ref{pureground}).

\subsubsection{Separable and entangled QGIFs}

Considering $\{|\underline{a}>_{\mathcal{A}};\underline{a}=1,2,...k_{a}\}\in
\mathcal{H}_{\mathcal{A}}$ and $\{|\underline{b}>_{\mathcal{B}};\underline{b}%
=1,2,...k_{b}\}\in \mathcal{H}_{\mathcal{B}}$ as orthonormal bases, we can
parameterize a pure total ground state in the form%
\begin{equation}
|\Psi >=\sum_{\underline{a}\underline{b}}c_{\underline{a}\underline{b}}|%
\underline{a}>_{\mathcal{A}}\otimes |\underline{b}>_{\mathcal{B}},
\label{groundstate}
\end{equation}%
where $c_{\underline{a}\underline{b}}$ is a complex matrix of dimension $%
\dim \mathcal{H}_{\mathcal{A}}\times \dim \mathcal{H}_{\mathcal{B}}.$ When
such coefficients factorize, $c_{\underline{a}\underline{b}}=c_{\underline{a}%
}c_{\underline{b}},$ we obtain a separable ground state (equivalently, pure
product state), when
\begin{equation*}
|\Psi >=|\Psi _{\mathcal{A}}>\otimes |\Psi _{\mathcal{B}}>,\mbox{ for }|\Psi
_{\mathcal{A}}>=\sum_{\underline{a}}c_{\underline{a}}|\underline{a}>_{%
\mathcal{A}}\mbox{ and }|\Psi _{\mathcal{B}}>=\sum_{\underline{b}}c_{%
\underline{b}}|\underline{b}>_{\mathcal{B}}.
\end{equation*}%
The entanglement entropy $\ _{q}^{\shortmid }\mathcal{S}(\ ^{\shortmid}\rho
_{\mathcal{A}})=0$ if and only if the pure ground state is separable. For
QGIFs, such definitions are motivated because corresponding sub-systems are
described by corresponding effective relativistic Hamilton functions, $%
\widetilde{H}_{\mathcal{A}}$ and $\widetilde{H}_{\mathcal{B}},$ and/or
effective thermodynamics energies, $\ _{\mathcal{A}}^{\shortmid }\mathcal{E}$
and $\ _{\mathcal{B}}^{\shortmid }\mathcal{E}.$

A ground state $|\Psi >$ (\ref{groundstate}) is \textit{entangled
(inseparable )} if $c_{\underline{a}\underline{b}}\neq c_{\underline{a}}c_{%
\underline{b}}$. For such a state, the entanglement entropy is positive, $\
_{q}^{\shortmid }\mathcal{S}(\ ^{\shortmid }\rho _{\mathcal{A}})>0.$ Using
quantum Schmidt decompositions (\ref{schm1}) and (\ref{schm2}), we prove
that
\begin{equation}
\ _{q}^{\shortmid }\mathcal{S=-}\sum_{\underline{a}}^{\min (\underline{a},%
\underline{b})}p_{\underline{a}}\log p_{\underline{a}}\mbox{ and }\
_{q}^{\shortmid }\mathcal{S}_{|\max }=\log \min (\underline{a},\underline{b})%
\mbox{ for }\sum_{\underline{a}}p_{\underline{a}}=1\mbox{ and }p_{\underline{%
a}}=1/\min (\underline{a},\underline{b}),\forall a.  \label{aux04}
\end{equation}

In summary, an entangled state of QGIFs is a superposition of several
quantum states associated to GIFs. An observer having access only to a
subsystem $\mathcal{A}$ will find him/ herself in a mixed state when the
total ground state $|\Psi >$ is entangled following such conditions:%
\begin{equation*}
\begin{array}{ccc}
|\Psi >:\mbox{ separable } & \longleftrightarrow & \ ^{\shortmid }\rho _{%
\mathcal{A}}:\mbox{ pure state}, \\
|\Psi >:\mbox{ entangled } & \longleftrightarrow & \ ^{\shortmid }\rho _{%
\mathcal{A}}:\mbox{ mixed state}.%
\end{array}%
\end{equation*}%
The von Neumann entanglement entropy $\ _{q}^{\shortmid }\mathcal{S}$
encodes two types of information: 1) how geometric evolution is quantum flow
correlated and 2) how much a given QGIF state differs from a separable QM
state. A maximum value of quantum correlations is reached when a given QGIF
state is a superposition of all possible quantum states with an equal
weight. Additional GIF properties are characterized by W-entropy $\
^{\shortmid }\mathcal{W}$ (\ref{wfperelmctl}) and thermodynamic entropy $\
^{\shortmid }\mathcal{S}$ (\ref{8rdthvhs}) which can be computed in certain
quasi-classical QM limits, for a 3+1 splitting, for instance, along a time
like curve.

\subsubsection{Two QGIFs systems as analogs of two spin and/or bipartite
systems}

The most simple example of an entangled system \cite{preskill,witten18,aolita14,nishioka18} is that of two particles $A$ and $B$ with spin $1/2$. In the information theory, such quantum spin systems can be
used to encode binary information as bits and, with further generalizations, to elaborate on quantum bits, qubits. Respective theoretical descriptions use density matrices and the von Neumann entropy.

To study similar entanglement properties of geometric flows in classical and quantum information theory we can consider two thermodynamical models of general covariant mechanical systems $\mathcal{A}=\left[ \mathbf{g,}\ ^{\shortmid }\mathcal{E},\ ^{\shortmid }\mathcal{S},\ ^{\shortmid }\eta %
\right] $ and $\mathcal{B}=\left[ \ _{1}\mathbf{g,}\ _{1}^{\shortmid }%
\mathcal{E},\ _{1}^{\shortmid }\mathcal{S},\ _{1}^{\shortmid }\eta \right] ,$
see formulas (\ref{8rdthvhs}). A respective QGIF model with entanglement is elaborated for different associated relativistic Hamiltonians and respective d-metrics $\mathbf{g}$ and$\ _{1}\mathbf{g}.$ For simplicity, we consider that the conventional Hilbert spaces are spanned by two orthonormal basic
states in the form $\{|\underline{a}>_{\mathcal{A}};\underline{a}=1,2\}\in \mathcal{H}_{\mathcal{A}}$ and $\{|\underline{b}>_{\mathcal{B}};\underline{b} =1,2\}\in \mathcal{H}_{\mathcal{B}},$ when $_{\mathcal{A},\mathcal{B}}<\underline{a}|\underline{b}>_{\mathcal{A},\mathcal{B}}=\delta _{\underline{a}\underline{b}}.$  The total Hilbert space $\mathcal{H}_{\mathcal{AB}}=%
\mathcal{H}_{\mathcal{A}}\otimes \mathcal{H}_{\mathcal{B}}$ has a 4-dim
orthonormal basis $\mathcal{H}_{\mathcal{AB}}=\{|11>,|12>,|21>,|22>\},$ where
$|\underline{a}\underline{b}>=|\underline{a}>_{\mathcal{A}}\otimes |%
\underline{b}>_{\mathcal{B}}$ are tensor product states.

As a general state, we can consider
\begin{equation}
|\Psi >=\cos \theta |12>-\sin \theta |21>,  \label{2flow}
\end{equation}%
where $0\leq \theta \leq \pi /2.$ The corresponding entanglement entropy (%
\ref{entangentr}) is computed%
\begin{equation*}
\ _{q}^{\shortmid }\mathcal{S}(\ ^{\shortmid }\rho _{\mathcal{A}})=-\cos
^{2}\theta \log (\cos ^{2}\theta )-\sin ^{2}\theta \log (\sin ^{2}\theta ).
\end{equation*}

Above formulas show that for $\theta =0,\pi /2$ we obtain pure product
states with zero entanglement entropy. For a system $|\Psi >=\frac{1}{\sqrt{2%
}}(|12>-|21>),$ when the density matrix
\begin{equation*}
\ ^{\shortmid }\rho _{\mathcal{A}}=\frac{1}{2}(|1>_{\mathcal{A}}\ _{\mathcal{%
A}}<1|+|2>_{\mathcal{A}}\ _{\mathcal{A}}<2|)=\frac{1}{2}diag(1,1)
\end{equation*}%
results in
\begin{equation*}
\ _{q}^{\shortmid }\mathcal{S}(\ ^{\shortmid }\rho _{\mathcal{A}})=-tr_{%
\mathcal{A}}(\ ^{\shortmid }\rho _{\mathcal{A}}\log \ ^{\shortmid }\rho _{%
\mathcal{A}})=\log 2.
\end{equation*}
So, the maximal entanglement is for $\theta =\pi /4.$ If the GIF structure
is "ignored" for such a quantum system (or (\ref{2flow})), we can treat it
as conventional QM system, for instance, with up-spin $|1>$ and down-spin $%
|2>.$ In a general context, QGIFs with nonholonomic structure determined by
Hamilton mechanical systems are characterized additionally by respective
values of W-entropy $\ ^{\shortmid }\mathcal{W}$ (\ref{wfperelmctl}) and
thermodynamic entropy $\ ^{\shortmid }\mathcal{S}$ (\ref{8rdthvhs}). In
orthonormal quantum bases, the entanglement entropy is the measure of "pure"
quantum entanglement. The information flows with rich nonholonomic geometric
structure are characterized additionally by geometric type entropies.

\subsubsection{Thermofield double QGIF states and entanglement and W-entropy}

If the evolution parameter $\beta =T^{-1}$ is treated as a temperature one
like in the standard G. Perelman's approach, we can consider respective
geometric flow theories as certan classical and/or quantum thermofield
models. Such a nontrivial example with entanglement and a thermofield double
GIFs state is defined by a ground state (\ref{groundstate}) parameterized in
the form
\begin{equation}
|\Psi >=Z^{-1/2}\sum\limits_{\underline{k}}e^{-\beta E_{\underline{k}}/2}|%
\underline{k}>_{\mathcal{A}}\otimes |\underline{k}>_{\mathcal{B}},
\label{dtfst}
\end{equation}%
where the normalization of the states is take for the partition function $%
Z=\sum\limits_{\underline{k}}e^{-\beta E_{\underline{k}}/2}$. Such values
are associated to the thermodynamic generating function $\ ^{\shortmid }%
\mathcal{Z}[\ ^{\shortmid }\mathbf{g}(\tau )]$ (\ref{genarthermfunct}) and
state density matrix $\ ^{\shortmid }\sigma (\beta ,\ ^{\shortmid }\mathcal{E%
}\ ,\ ^{\shortmid }\mathbf{g})$ (\ref{statedens}) the energy $\ ^{\shortmid }%
\mathcal{E}_{\mathcal{A}}=\{E_{\underline{k}}\}\ $\ is considered quantized
with a discrete spectrum for a QGIF system $\mathcal{A}=\left[ \mathbf{g,}\
^{\shortmid }\mathcal{E},\ ^{\shortmid }\mathcal{S},\ ^{\shortmid }\eta %
\right] .$ The density matrix for this subsystem determining a Gibbse state
is computed \
\begin{equation*}
\ ^{\shortmid }\rho _{\mathcal{A}}=Z^{-1}\sum\limits_{\underline{k}%
}e^{-\beta E_{\underline{k}}/2}|\underline{k}>_{\mathcal{A}}\otimes \ _{%
\mathcal{A}}<\underline{k}|=Z^{-1}e^{-\beta \ ^{\shortmid }\mathcal{E}_{%
\mathcal{A}}}.
\end{equation*}%
In above formulas, we consider $\ ^{\shortmid }\mathcal{E}_{\mathcal{A}}$ as
\ a (modular) Hamiltonian $\ ^{\shortmid }\mathcal{E}_{\mathcal{A}}$ such
that $\ ^{\shortmid }\mathcal{E}_{\mathcal{A}}|\underline{k}>_{\mathcal{A}%
}=E_{\underline{k}}|\underline{k}>_{\mathcal{A}}$.

In principle, the thermofield double states for QGIFs consist certain
entanglement purifications of thermal states with Boltzman weight $%
p_{k}=Z^{-1}\sum\limits_{\underline{k}}e^{-\beta E_{\underline{k}}}$, see
discussions related to formulas (\ref{aux03}). Coping the state vectros $\{|%
\underline{k}>_{\mathcal{B}}\}$ from $\mathcal{H}_{\mathcal{A}}$ to $%
\mathcal{H}_{\mathcal{B}},$ we can purify the QGIF thermal system $\mathcal{A%
}$ in the extended Hilbert space $\mathcal{H}_{\mathcal{A}}$ $\otimes $ $%
\mathcal{H}_{\mathcal{B}}.$ In result, every expectation value of local
operators in $\mathcal{A}$ can be represented using the thermofield double
state $|\Psi >$ (\ref{dtfst}) of the total system $\mathcal{A\cup B}$. \ For
such models, the entanglement entropy is a measure of the thermal entropy of
the subsystem $\mathcal{A}$ when
\begin{equation*}
\ ^{\shortmid }\mathcal{S}(\ ^{\shortmid }\rho _{\mathcal{A}})=-tr_{\mathcal{%
A}}[\ ^{\shortmid }\rho _{\mathcal{A}}(-\beta \ ^{\shortmid }\mathcal{E}_{%
\mathcal{A}}-\log Z)]=\beta (<\ ^{\shortmid }\mathcal{E}_{\mathcal{A}}>-\
^{\shortmid }\mathcal{F}_{\mathcal{A}}),
\end{equation*}%
where the thermal free energy is computed $\ ^{\shortmid }\mathcal{F}_{%
\mathcal{A}}=-\log Z.$ Here we note that for the thermofield values it is
omitted the label "q" considered, for instance, for $\ _{q}^{\shortmid }%
\mathcal{S}$ (\ref{entangentr}), see also formulas (\ref{condhamentr}).

Thermofield GIF configurations are also characterized by the respective
W-entropy $\ ^{\shortmid }\mathcal{W}$ (\ref{wfperelmctl}) which can be
defined even thermodynamic models are not elaborated. For nonholonomic
kinetic, diffusion and thermodynamic structures including relativistic Ricci
flows, such models were studied in detail in \cite%
{vacaru00ap,ruchin13,vdiffusion}, see references therein. We also cite some
important works on geometric thermodynamics and thermofield theories, see
\cite{ruppeiner,quevedo,castro19} and references. The thermofield double
states were considered in black hole thermodynamics and QFT, see reviews of
results in \cite{preskill,solodukhin11,witten18,aolita14,nishioka18}.

\subsubsection{Bell like QGIF states}

In a two QGIF system, a \ state (\ref{2flow}) is maximally entangled for $%
\theta =\pi /4$. Analogs of Bell state (or Einstein-Podosly-Rosen \ pairs)
in quantum geometric flow theory are defined
\begin{eqnarray}
|\Psi _{\mathcal{B}}^{1} >&=&\frac{1}{\sqrt{2}}(|11>+|22>),|\Psi _{\mathcal{B%
}}^{2}>=\frac{1}{\sqrt{2}}(|11>-|22>),  \label{bellqgif} \\
|\Psi _{\mathcal{B}}^{3} >&=&\frac{1}{\sqrt{2}}(|12>+|21>),|\Psi _{\mathcal{B%
}}^{4}>=\frac{1}{\sqrt{2}}(|12>-|21>).  \notag
\end{eqnarray}%
In QM models, these states violate the Bell's inequalities. Such
inequalities hold in a hidden variable theory for the probabilistic features
of QM with a hidden variable and a probability density. In this work, the
states (\ref{bellqgif}) encode also information of geometric flows
characterized by W-entropy.

\paragraph{EPR pairs and multi-qubits for QGIFs:}

The constructions can be extended for systems of $k$ quabits. The first
example generalizes the concept of Greenberger-Horne-Zelinger, GHZ, states
\cite{greenberger89,greenberger90,nishioka18},
\begin{equation*}
|\Psi _{\mathcal{B}}^{GHZ}>=\frac{1}{\sqrt{2}}(|1>^{\otimes k}+|2>^{\otimes
k}).
\end{equation*}%
In quantum information theory, thre are used another type of entangled
states (called W states; do not confuse with W-entropy) \cite{dur2000},%
\begin{equation*}
|\Psi _{\mathcal{B}}^{W}>=\frac{1}{\sqrt{2}}%
(|21...11>+|121...1>+...+|11...12>).
\end{equation*}%
We emphasize that $|\Psi _{\mathcal{B}}^{GHZ}>$ $\ $is fully separable but
not $|\Psi _{\mathcal{B}}^{W}>$ which we shall prove in the example below.

\paragraph{Tripartite QGIFs:}

For $k=3$ with subsystems $\mathcal{A},\mathcal{B}$ and $\mathcal{C},$ we
write
\begin{equation*}
|\Psi _{\mathcal{B}}^{GHZ}>=\frac{1}{\sqrt{2}}(|111>+|222>)\mbox{ and }|\Psi
_{\mathcal{B}}^{W}>=\frac{1}{\sqrt{2}}(|112>+|121>+|211>).
\end{equation*}%
Considering $Tr_{\mathcal{C}},$ we define \ the reduced density matrices for
the system $\mathcal{A\cup B},$%
\begin{equation*}
\ ^{\shortmid }\rho _{\mathcal{A\cup B}}^{GHZ}=\frac{1}{2}(|11><11|+|22><22|)%
\mbox{ and }\ ^{\shortmid }\rho _{\mathcal{A\cup B}}^{W}=\frac{2}{3}|\Psi _{%
\mathcal{B}}^{3}><\Psi _{\mathcal{B}}^{3}|+\frac{1}{3}|11><11|.
\end{equation*}%
This describe two different QGIF states. The first one is fully separable
and can be represented in the form $\ ^{\shortmid }\rho _{\mathcal{A\cup B}%
}^{GHZ}=\sum\limits_{\underline{k}=1}^{2}p_{\underline{k}}\ ^{\shortmid
}\rho _{\mathcal{A}}^{\underline{k}}\otimes \ ^{\shortmid }\rho _{\mathcal{B}%
}^{\underline{k}},$ where $p_{\underline{k}}=1/2$ and $\ ^{\shortmid }\rho _{%
\mathcal{A},\mathcal{B}}^{\underline{1}}=|11><11|$ and $\ ^{\shortmid }\rho
_{\mathcal{A},\mathcal{B}}^{\underline{2}}=|22><22|.$ Because of the Bell
state $|\Psi _{\mathcal{B}}^{3}>$ (\ref{bellqgif}), $\ $the $\ ^{\shortmid
}\rho _{\mathcal{A\cup B}}^{W}$ can not be written in a separable form. So,
the state $|\Psi _{\mathcal{B}}^{W}>$ is still entangled even we have taken $%
Tr_{\mathcal{C}}.$ This establishes a quantum correlation between QGIFs.
Additionally, such values are characterized by W-entropies of type $\
^{\shortmid }\mathcal{W}$ (\ref{wfperelmctl}) computed for $\mathcal{A},%
\mathcal{B},\mathcal{C}$ and $\mathcal{A\cup B}.$

\subsection{Important properties and entanglement inequalities for QGIFs
entropies}

We summarize several useful properties of the entanglement entropy (\ref%
{entangentr}) for QGIFs formulated in terms of the density matrix of type $\
^{\shortmid }\rho _{\mathcal{A}}=Tr_{\mathcal{H}_{\mathcal{B}}}(\
_{t}^{\shortmid }\rho ).$ We omit explicit cumbersome and techniqual proofs
because they are similar to derivations in \cite{nielsen10}. For any $\
^{\shortmid }\rho _{\mathcal{A}}$  associated to a state density $\
^{\shortmid }\rho (\beta ,\ ^{\shortmid }\mathcal{E}\ ,\ ^{\shortmid }%
\mathbf{g})$ of type (\ref{statedens}), we can compute the respective W-entropy
and geometric thermodynamic entropy taking measures determined by $\
^{\shortmid }\mathbf{g}$ and/or respective Hamilton mechanical variables.
Rigorous mathematical proofs involve a geometric analysis technique summarized
in \cite{perelman1,monogrrf1,monogrrf2,monogrrf3}. For applications in
modern gravity and particle physics theories, we can elaborate on
alternative approaches using the anholonomic frame method of constructing
off-diagonal solutions in relativistic geometric flow theories and
generalizations \cite{ruchin13,gheorghiu16,bubuianu19}. Using explicit
classes of solutions and re-defining normalizing functions, we can always
compute Perelman's like entropy functionals at least in the quasi-classical
limit with respective measures and related to $\ _{q}^{\shortmid }\mathcal{S}
$ (\ref{entangentr}) for a QGIF or a thermofield GIF model.

\subsubsection{(Strong) subadditivity}

We present four important properties of QGIFs which result in the strong
subadditivity property of entanglement and Perelman's entropies.

\paragraph{Entanglement entropy for complementary subsystems:}

If $\mathcal{B}=\overline{\mathcal{A}},$ the entanglement entropies are the same%
\begin{equation*}
\ _{q}^{\shortmid }\mathcal{S}_{\mathcal{A}}=\ _{q}^{\shortmid }\mathcal{S}_{%
\overline{\mathcal{A}}}
\end{equation*}%
which follows from formulas (\ref{aux04}) for a pure ground state wave
function. Similar equalities for the W-entropy $\ ^{\shortmid }\mathcal{W}$ (%
\ref{wfperelmctl}) and/or thermodynamic entropy $\ ^{\shortmid }\mathcal{S}$
(\ref{8rdthvhs}) can be proven only for the same d-metrics $\ ^{\shortmid }%
\mathbf{g}$ and respective normalizations on $\mathcal{A}$ and $\overline{%
\mathcal{A}}.$ Here we note that $\ _{q}^{\shortmid }\mathcal{S}_{\mathcal{A}%
}\neq \ _{q}^{\shortmid }\mathcal{S}_{\mathcal{B}}$ if $\mathcal{A\cup B}$
is a mixed state, for instance, at a finite temperature. So, in general,
\begin{equation*}
\ _{q}^{\shortmid }\mathcal{S}_{\mathcal{A}}\neq \ _{q}^{\shortmid }\mathcal{%
S}_{\mathcal{B}}\mbox{ and }\ _{q}^{\shortmid }\mathcal{W}_{\mathcal{A}}\neq
\ _{q}^{\shortmid }\mathcal{W}_{\mathcal{B}}.
\end{equation*}%
We have to consider a subclass of nonholonomic deformations when conditions transform into equalities for respective relativistic flow evolution scenarios and associated thermodynamic and QM systems.

\paragraph{Subadditivity:}

For \ disjoint subsystems $\mathcal{A}$ and $\mathcal{B}$, there are satisfied the conditions of subadditivity
\begin{equation}
\ _{q}^{\shortmid }\mathcal{S}_{\mathcal{A\cup B}}\leq \ _{q}^{\shortmid }%
\mathcal{S}_{\mathcal{A}}+\ _{q}^{\shortmid }\mathcal{S}_{\mathcal{B}}%
\mbox{
and }|\ _{q}^{\shortmid }\mathcal{S}_{\mathcal{A}}-\ _{q}^{\shortmid }%
\mathcal{S}_{\mathcal{B}}|\leq \ _{q}^{\shortmid }\mathcal{S}_{\mathcal{%
A\cup B}}.  \label{subaditcond}
\end{equation}%
The second equation transforms into the triangle inequality \cite{araki70}. In the quasi-classical limit, we obtain similar inequalities for the thermodynamic entropy $\ ^{\shortmid }\mathcal{S}$ (\ref{8rdthvhs}). We
claim that similar conditions hold for the W-entropy $\ ^{\shortmid }\mathcal{W}$ (\ref{wfperelmctl}). They can be  computed as quantum perturbations in a QFT associated to a bipartite QGIF model%
\begin{equation*}
\ _{q}^{\shortmid }\mathcal{W}_{\mathcal{A\cup B}}\leq \ _{q}^{\shortmid }%
\mathcal{W}_{\mathcal{A}}+\ _{q}^{\shortmid }\mathcal{W}_{\mathcal{B}}%
\mbox{
and }|\ _{q}^{\shortmid }\mathcal{W}_{\mathcal{A}}-\ _{q}^{\shortmid }%
\mathcal{W}_{\mathcal{B}}|\leq \ _{q}^{\shortmid }\mathcal{W}_{\mathcal{%
A\cup B}}.
\end{equation*}%
Such flow evolution and QM scenarios are elaborated for mixed geometric and
quantum probabilistic information flows.

\paragraph{Strong subadditivity:}

Considering three disjointed QGIF subsystems $\mathcal{A},\mathcal{B}$ and $%
\mathcal{C}$ and certain conditions of convexity of a function built from
respective density matrix and unitarity of systems \cite%
{lieb73,narnhofer85,witten18,nishioka18}, one hold the following
inequalities of \textit{strong subbadditivity}:%
\begin{equation*}
\ _{q}^{\shortmid }\mathcal{S}_{\mathcal{A\cup B\cup C}}+\ _{q}^{\shortmid }%
\mathcal{S}_{\mathcal{B}}\leq \ \ _{q}^{\shortmid }\mathcal{S}_{\mathcal{%
A\cup B}}+\ _{q}^{\shortmid }\mathcal{S}_{B\mathcal{\cup C}}\mbox{ and }\
_{q}^{\shortmid }\mathcal{S}_{\mathcal{A}}+\ _{q}^{\shortmid }\mathcal{S}_{%
\mathcal{C}}\leq \ _{q}^{\shortmid }\mathcal{S}_{\mathcal{A\cup B}}+\
_{q}^{\shortmid }\mathcal{S}_{B\mathcal{\cup C}}.
\end{equation*}%
From these conditions, the conditions of subadditivity (\ref{subaditcond})
can be derived as particular cases. Along causal curves on respective
cotangent Lorentz manifolds, we can prove similar formulas for the W-entropy
and small quantum perturbations%
\begin{equation*}
\ _{q}^{\shortmid }\mathcal{W}_{\mathcal{A\cup B\cup C}}+\ _{q}^{\shortmid }%
\mathcal{W}_{\mathcal{B}}\leq \ \ _{q}^{\shortmid }\mathcal{W}_{\mathcal{%
A\cup B}}+\ _{q}^{\shortmid }\mathcal{W}_{B\mathcal{\cup C}}\mbox{ and }\
_{q}^{\shortmid }\mathcal{W}_{\mathcal{A}}+\ _{q}^{\shortmid }\mathcal{W}_{%
\mathcal{C}}\leq \ _{q}^{\shortmid }\mathcal{W}_{\mathcal{A\cup B}}+\
_{q}^{\shortmid }\mathcal{W}_{B\mathcal{\cup C}}.
\end{equation*}%
We claim such properties for respective QGIFs. They play vital roles in the
entropic proofs of the so-called $c$- $\ F$-theorems for renormalization
group flows in QFT, see review of results in section VIII of \cite%
{nishioka18}. In our approach, we elaborate on a different geometric
formalism with nonholonomic flow evolution and respective applications in
quantum information theory.

\subsubsection{Relative entropy and QGIF entanglement}

There are several measures of quantum entanglement which are determined by
geometric and thermodynamic values for QGIFs. We begin with the concept of%
\textit{\ relative entropy} in geometric information theories.
\begin{equation}
\ ^{\shortmid }\mathcal{S}(\ ^{\shortmid }\rho _{\mathcal{A}}\shortparallel
\ ^{\shortmid }\sigma _{\mathcal{A}})=Tr_{\mathcal{H}_{\mathcal{B}}}[\
^{\shortmid }\rho _{\mathcal{A}}(\log \ ^{\shortmid }\rho _{\mathcal{A}%
}-\log \ ^{\shortmid }\sigma _{\mathcal{A}})],  \label{relativentr}
\end{equation}%
where $\ ^{\shortmid }\mathcal{S}(\ ^{\shortmid }\rho _{\mathcal{A}%
}\shortparallel \ ^{\shortmid }\rho _{\mathcal{A}})=0.$ This value is a
measure of "distance" between two QGIFs with a norm $||\ ^{\shortmid }\rho _{%
\mathcal{A}}||=tr(\sqrt{(\ ^{\shortmid }\rho _{\mathcal{A}})(\ ^{\shortmid
}\rho _{\mathcal{A}}^{\dag })}).$ For thermodynamical GIF systems, it
transforms into the conditional entropy (\ref{condhamentr}). It was
introduced and studied for standard densiti matrices in QM and information
theory, respectively, in \cite{umegaki62} and \cite{vedral02,ohya04}, see
reviews \cite{preskill,witten18,nishioka18}. In straightforward form, we can
check that there are satisfied certain important properties and inequalities.

\paragraph{ Two QGIF systems}

are characterized by formulas and conditions:
\begin{enumerate}
\item for tensor products of density matrices,
\begin{equation*}
\ ^{\shortmid }\mathcal{S}(\ _{1}^{\shortmid }\rho _{\mathcal{A}}\otimes \
_{2}^{\shortmid }\rho _{\mathcal{A}}\shortparallel \ _{1}^{\shortmid }\sigma
_{\mathcal{A}}\otimes \ _{2}^{\shortmid }\sigma _{\mathcal{A}})=\
^{\shortmid }\mathcal{S}(\ _{1}^{\shortmid }\rho _{\mathcal{A}%
}\shortparallel \ _{1}^{\shortmid }\sigma _{\mathcal{A}})+\ ^{\shortmid }%
\mathcal{S}(\ _{2}^{\shortmid }\rho _{\mathcal{A}}\shortparallel \
_{2}^{\shortmid }\sigma _{\mathcal{A}});
\end{equation*}

\item positivity:%
\begin{equation*}
\ ^{\shortmid }\mathcal{S}(\ ^{\shortmid }\rho _{\mathcal{A}}\shortparallel
\ ^{\shortmid }\sigma _{\mathcal{A}})\geq \frac{1}{2}||\ ^{\shortmid }\rho _{%
\mathcal{A}}-\ ^{\shortmid }\sigma _{\mathcal{A}}||^{2},\mbox{ i.e. }\
^{\shortmid }\mathcal{S}(\ ^{\shortmid }\rho _{\mathcal{A}}\shortparallel \
^{\shortmid }\sigma _{\mathcal{A}})\geq 0;
\end{equation*}

\item monotonicity:%
\begin{equation*}
\ ^{\shortmid }\mathcal{S}(\ ^{\shortmid }\rho _{\mathcal{A}}\shortparallel
\ ^{\shortmid }\sigma _{\mathcal{A}})\geq \ ^{\shortmid }\mathcal{S}(tr_{s}\
^{\shortmid }\rho _{\mathcal{A}}|tr_{s}\ ^{\shortmid }\sigma _{\mathcal{A}}),
\end{equation*}%
where $tr_{s}$ is the trase for a subsystem of $\mathcal{A}.$
\end{enumerate}

Using above positivity formula and the Schwarz inequality $||X||\geq
tr(XY)/||X||,$ we obtain that
\begin{equation*}
\ ^{\shortmid }\mathcal{S}(\ ^{\shortmid }\rho _{\mathcal{A}}\shortparallel
\ ^{\shortmid }\sigma _{\mathcal{A}})\geq \frac{1}{2}\frac{(\langle \mathcal{%
O}\rangle _{\rho }-\langle \mathcal{O}\rangle _{\sigma })^{2}}{||\mathcal{O}%
||^{2}}
\end{equation*}%
for any expectation value $\langle \mathcal{O}\rangle _{\rho }$ of an
operator $\mathcal{O}$ computed with the density matrix $\ ^{\shortmid }\rho
_{\mathcal{A}},$ see formulas (\ref{expectvalues}).$\,$

The relative entropy $\ ^{\shortmid }\mathcal{S}(\ ^{\shortmid }\rho _{%
\mathcal{A}}\shortparallel \ ^{\shortmid }\sigma _{\mathcal{A}})$ (\ref%
{relativentr}) can be related to the entaglement entropy $\ _{q}^{\shortmid }%
\mathcal{S}(\ ^{\shortmid }\rho _{\mathcal{A}})$ (\ref{entangentr}) using
formula%
\begin{equation}
\ ^{\shortmid }\mathcal{S}(\ ^{\shortmid }\rho _{\mathcal{A}}\shortparallel
1_{\mathcal{A}}/k_{\mathcal{A}})=\log k_{\mathcal{A}}-\ _{q}^{\shortmid }%
\mathcal{S}(\ ^{\shortmid }\rho _{\mathcal{A}}),  \label{aux06}
\end{equation}%
where $1_{\mathcal{A}}$ is the $k_{\mathcal{A}}\times k_{\mathcal{A}}$ unit
matrix for a $k_{\mathcal{A}}$-dimensional Hilbert space associated to the
region $\mathcal{A}.$ Above properties can be re-defined by the entanglement
entropy $\ _{q}^{\shortmid }\mathcal{S}$, see similar formulas for QGIFs in
Hamilton mechancal variables in \cite{vacaru19b}.

\paragraph{ Three QGIF systems:}

Let us denote by $\ ^{\shortmid }\rho _{\mathcal{A\cup B\cup C}}$ the
density matrix of three QGIFs subsystems $\mathcal{A\cup B\cup C}$ and, for
instance, $\ ^{\shortmid }\rho _{\mathcal{A\cup B}}$ for its restriction on $%
\mathcal{A\cup B}$ and $\ ^{\shortmid }\rho _{\mathcal{B}}$ for its
restriction on $\mathcal{B}.$ Using the formula for computing traces of
reduced density matrices,%
\begin{equation*}
tr_{\mathcal{A\cup B\cup C}}[\ ^{\shortmid }\rho _{\mathcal{A\cup B\cup C}}(%
\mathcal{O}_{\mathcal{A\cup B}}\otimes 1_{\mathcal{C}}/k_{\mathcal{C}})]=tr_{%
\mathcal{A\cup B}}(\ ^{\shortmid }\rho _{\mathcal{A\cup B}}\mathcal{O}_{%
\mathcal{A\cup B}})
\end{equation*}%
we prove such identities
\begin{eqnarray*}
\ ^{\shortmid }\mathcal{S}(\ ^{\shortmid }\rho _{\mathcal{A\cup B\cup C}}
\shortparallel 1_{\mathcal{A\cup B\cup C}}/k_{\mathcal{A\cup B\cup C}})&=&\
^{\shortmid }\mathcal{S}(\ ^{\shortmid }\rho _{\mathcal{A\cup B}%
}\shortparallel 1_{\mathcal{A\cup B}}/k_{\mathcal{A\cup B}})+\ ^{\shortmid }%
\mathcal{S}(\ ^{\shortmid }\rho _{\mathcal{A\cup B\cup C}}\shortparallel \
^{\shortmid }\rho _{\mathcal{A\cup B}}\otimes 1_{\mathcal{C}}/k_{\mathcal{C}%
}), \\
\ ^{\shortmid }\mathcal{S}(\ ^{\shortmid }\rho _{\mathcal{B\cup C}}
\shortparallel 1_{\mathcal{B\cup C}}/k_{\mathcal{B\cup C}})&=&\ ^{\shortmid }%
\mathcal{S}(\ ^{\shortmid }\rho _{\mathcal{B}}\shortparallel 1_{\mathcal{B}%
}/k_{\mathcal{B}})+\ ^{\shortmid }\mathcal{S}(\ ^{\shortmid }\rho _{\mathcal{%
B\cup C}}\shortparallel \ ^{\shortmid }\rho _{\mathcal{B}}\otimes 1_{%
\mathcal{C}}/k_{\mathcal{C}});
\end{eqnarray*}%
and inequalities {\small
\begin{eqnarray*}
&&\ ^{\shortmid }\mathcal{S}(\ ^{\shortmid }\rho _{\mathcal{A\cup B\cup C}}
\shortparallel \ ^{\shortmid }\rho _{\mathcal{A\cup B}}\otimes 1_{\mathcal{C}%
}/k_{\mathcal{C}})\geq \ ^{\shortmid }\mathcal{S}(\ ^{\shortmid }\rho _{%
\mathcal{B\cup C}}\shortparallel \ ^{\shortmid }\rho _{\mathcal{B}}\otimes
1_{\mathcal{C}}/k_{\mathcal{C}}), \\
&&\ ^{\shortmid }\mathcal{S}(\ ^{\shortmid }\rho _{\mathcal{A\cup B\cup C}}
\shortparallel 1_{\mathcal{A\cup B\cup C}}/k_{\mathcal{A\cup B\cup C}})+\
^{\shortmid }\mathcal{S}(\ ^{\shortmid }\rho _{\mathcal{B}}\shortparallel 1_{%
\mathcal{B}}/k_{\mathcal{B}})\geq \ ^{\shortmid }\mathcal{S}(\ ^{\shortmid
}\rho _{\mathcal{A\cup B}} \shortparallel 1_{\mathcal{A\cup B}}/k_{\mathcal{%
A\cup B}})+\ ^{\shortmid }\mathcal{S}(\ ^{\shortmid }\rho _{\mathcal{B\cup C}%
}\shortparallel 1_{\mathcal{B\cup C}}/k_{\mathcal{B\cup C}}).
\end{eqnarray*}%
} These formulas can be re-written (after corresponding applications of the rule (\ref%
{aux06})) for the entanglement entropies $\ _{q}^{\shortmid }\mathcal{S}$
and Hamilton mechanical variables with "tilde" \cite{vacaru19b}.

\subsubsection{Mutual information for QGIFs}

The correlation between two QGIF systems $\mathcal{A}$ and $\mathcal{B}$ (it
can be involved also a third system $\mathcal{C}$) is characterized by the
\textit{mutual information} $\ ^{\shortmid }\mathcal{J}(\mathcal{A},\mathcal{%
B})$ and respective inequalities which follow from above formulas for
relative entropy,%
\begin{equation*}
\ ^{\shortmid }\mathcal{J}(\mathcal{A},\mathcal{B}):=\ ^{\shortmid }\mathcal{%
S}_{\mathcal{A}}+\ ^{\shortmid }\mathcal{S}_{\mathcal{B}}-\ ^{\shortmid }%
\mathcal{S}_{\mathcal{A\cup B}}\geq 0\mbox{ and }\ ^{\shortmid }\mathcal{J}(%
\mathcal{A},\mathcal{B\cup C})\leq \ ^{\shortmid }\mathcal{J}(\mathcal{A},%
\mathcal{B}).
\end{equation*}%
The mutual information is related to the relative entropy following formula
\begin{equation}
\ ^{\shortmid }\mathcal{J}(\mathcal{A},\mathcal{B})=\ ^{\shortmid }\mathcal{S%
}(\ ^{\shortmid }\rho _{\mathcal{A\cup B}}\shortparallel \ ^{\shortmid }\rho
_{\mathcal{A}}\otimes \ ^{\shortmid }\rho _{\mathcal{B}}),
\label{mutualinf2}
\end{equation}%
which allows to consider similar concepts and inequalities for the
entanglement of QGIF systems:%
\begin{eqnarray*}
&&\ _{q}^{\shortmid }\mathcal{J}(\mathcal{A},\mathcal{B}):=\ _{q}^{\shortmid
}\mathcal{S}_{\mathcal{A}}+\ _{q}^{\shortmid }\mathcal{S}_{\mathcal{B}}-\
_{q}^{\shortmid }\mathcal{S}_{\mathcal{A\cup B}}\geq 0, \\
&&\ _{q}^{\shortmid }\mathcal{J}(\mathcal{A},\mathcal{B\cup C})\leq \
_{q}^{\shortmid }\mathcal{J}(\mathcal{A},\mathcal{B}),\mbox{ for }\
_{q}^{\shortmid }\mathcal{J}(\mathcal{A},\mathcal{B})=\ _{q}^{\shortmid }%
\mathcal{S}(\ ^{\shortmid }\rho _{\mathcal{A\cup B}}\shortparallel \
^{\shortmid }\rho _{\mathcal{A}}\otimes \ ^{\shortmid }\rho _{\mathcal{B}}).
\end{eqnarray*}

In the classical variant of GIFs, one hold similar formulas for GIFs and associated thermodynamic models with statistical density $\ ^{\shortmid}\rho (\beta ,\ ^{\shortmid }\mathcal{E}\ ,\ ^{\shortmid }\mathbf{g})$ (\ref{statedens}). For relativistic geometric flows, we claim that similar properties hold for the constructions using the W-entropy. In particular, this can be proven for causal configurations in nonholonomic Hamilton
variables \cite{vacaru19b}.

The mutual information between two QGIFs shows how much for an union $%
\mathcal{A\cup B}$ the density matrix $\ ^{\shortmid }\rho _{\mathcal{A\cup B%
}}$ differs from a separable state $\ ^{\shortmid }\rho _{\mathcal{A}%
}\otimes \ ^{\shortmid }\rho _{\mathcal{B}}.$ Quantum correlations entangle
even spacetime disconnected regions of the phase spacetime under geometric
flow evolution. For bounded operators $\mathcal{O}_{\mathcal{A}}$ and $%
\mathcal{O}_{\mathcal{B}}$ under geometric evolution in respective regions,
one holds true (the proof is similar to that in \cite{wolf08}) the inequality%
\begin{equation*}
\ ^{\shortmid }\mathcal{J}(\mathcal{A},\mathcal{B})\geq \frac{1}{2}\frac{%
(\langle \mathcal{O}_{\mathcal{A}}\mathcal{O}_{\mathcal{B}}\rangle -\langle
\mathcal{O}_{\mathcal{A}}\rangle \langle \mathcal{O}_{\mathcal{B}}\rangle
)^{2}}{||\mathcal{O}_{\mathcal{A}}||^{2}||\mathcal{O}_{\mathcal{B}}||^{2}}.
\end{equation*}%
Such formulas can be proven for associated thermodynamic systems to
classical GIFs using the statistical density if, for instance, $\mathcal{A}$
and $\mathcal{B}$ are certain subsystems of phase spaces and respective
geometric flows.

\subsubsection{The R\'{e}nyi entropy for QGIFs}

We can introduce another type of parametric entropy which provides us more
information about the eigenvalues of reduced entropy matrices thant the
entanglement entropy. This is the R\'{e}nyi entropy  \cite{renyi61}
which is important for computing the entanglement entropy of QFTs using the
replica method, see section IV of \cite{nishioka18}. Such constructions are
possible in QGIF theory because the thermodynamic generating function $\
^{\shortmid }\mathcal{Z}[\ ^{\shortmid }\mathbf{g}(\tau )]$ (\ref%
{genarthermfunct}) and related statistical density $^{\shortmid }\rho (\beta
,\ ^{\shortmid }\mathcal{E}\ ,\ ^{\shortmid }\mathbf{g})$ (\ref{statedens})
can be used for defining $\ ^{\shortmid }\sigma _{\mathcal{A}}$ (\ref{aux01}%
) \ as a probability distribution.

\paragraph{Replica method and G. Perelman's thermodynamica model:}

Let us consider an integer $r$ called as the replica parameter and introduce
the R\'{e}nyi entropy%
\begin{equation}
\ _{r}^{\shortmid }\mathcal{S}(\mathcal{A}):=\frac{1}{1-r}\log [tr_{\mathcal{%
A}}(\ ^{\shortmid }\rho _{\mathcal{A}})^{r}]  \label{renentr}
\end{equation}
for a QGIF\ system determined by a density matrix $\ ^{\shortmid }\rho _{%
\mathcal{A}}.$\footnote{%
We use the symbol $r$ for the replica parameter (and not $n$ as in the
typical works in information theory) because the symbol $n$ is used in our
works for the dimension of base manifolds.} We use the symbol $r$ for the
replica parameter (and not $n$ as in the typical works in information
theory) because the symbol $n$ is used in our works for the dimension of
base manifolds. To elaborate a computational formalism one considers an
analytic continuation of $r$ to a real number which allows us to define the
limit $\ _{q}^{\shortmid }\mathcal{S}(\ ^{\shortmid }\rho _{\mathcal{A}%
})=\lim_{r\rightarrow 1}\ _{r}^{\shortmid }\mathcal{S}(\mathcal{A})$,
 with the normalization $tr_{\mathcal{A}}(\ ^{\shortmid }\rho _{\mathcal{A}})$
for $r\rightarrow 1,$when the R\'{e}nyi entropy (\ref{renentr}) reduces to
the entanglement entropy (\ref{entangentr}).

There are satisfied certain important inequalities for derivatives on
replica parameter, $\partial _{r},$ of the R\'{e}nyi entropy $\
_{r}^{\shortmid }\mathcal{S}$ (proofs are similar to \cite{zycz03}):%
\begin{eqnarray}
\partial _{r}(\ _{r}^{\shortmid }\mathcal{S)} &\leq &0,  \label{aux07} \\
\partial _{r}\left( \frac{r-1}{r}\ _{r}^{\shortmid }\mathcal{S}\right) &\geq
&0,\partial _{r}[(r-1)\ _{r}^{\shortmid }\mathcal{S]}\geq 0,\partial
_{rr}^{2}[(r-1)](\ _{r}^{\shortmid }\mathcal{S)}\leq 0.  \notag
\end{eqnarray}%
These formulas have usual thermodynamical interpretations for a system with
a modular Hamiltonian $H_{\mathcal{A}}$ and effective statistical density $\
^{\shortmid }\rho _{\mathcal{A}}:=e^{-2\pi H_{\mathcal{A}}}.$ Considering $%
\beta _{r}=2\pi r$ as the inverse temperature, we introduce the effective
"thermal" statistical generation (partition) function,%
\begin{equation*}
\ _{r}^{\shortmid }\mathcal{Z}(\beta _{r}):=tr_{\mathcal{A}}(\ ^{\shortmid
}\rho _{\mathcal{A}})^{r}=tr_{\mathcal{A}}(e^{-\beta _{r}H_{\mathcal{A}}}).
\end{equation*}%
similarly to $\ ^{\shortmid }\mathcal{Z}[\ ^{\shortmid }\mathbf{g}(\tau )]$ (%
\ref{genarthermfunct}). In analogy to the thermodynamical model for
geometric flows (\ref{8rdthvhs}), we compute by canonical relations such
statistical mechanics values%
\begin{eqnarray*}
&&\ _{r}^{\shortmid }\mathcal{E}(\beta _{r}) :=-\partial _{\beta _{r}}\log
[\ _{r}^{\shortmid }\mathcal{Z}(\beta _{r})]\geq 0,%
\mbox{ for. the modular
energy}; \\
&&\ _{r}^{\shortmid }\mathcal{\breve{S}}(\beta _{r}) :=\left( 1-\beta
_{r}\partial _{\beta _{r}}\right) \log [\ _{r}^{\shortmid }\mathcal{Z}(\beta
_{r})]\geq 0,\mbox{ for. the modular entropy}; \\
&&\ _{r}^{\shortmid }\mathcal{C}(\beta _{r}) :=\beta _{r}^{2}\partial
_{\beta _{r}}^{2}\log [\ _{r}^{\shortmid }\mathcal{Z}(\beta _{r})]\geq 0,%
\mbox{ for. the modular capacity}.
\end{eqnarray*}%
These inequalities are equivalent to the second line in (\ref{aux07}) and
characterize the stability if GIFs as a thermal system with replica
parameter regarded as the inverse temperature for a respective modular
Hamiltonian. Such replica criteria of stability were not considered in the
original works on Ricci flows \cite{perelman1,monogrrf1,monogrrf2,monogrrf3}%
. They define a new direction for the theory of geometric flows and
applications in modern physics with respective genrealizations for
nonholonomic structures. \cite{vjmp08,vrmp09,vacaru00ap,vacaru09,rajpoot17,
ruchin13,gheorghiu16,vacaru19,vacaru19a,vacaru19b,vacaru18,bubuianu18}.

We note that the constructions with the modular entropy can be transformed
into models derived with the R\'{e}nyi entropy and inversely. Such
transforms can be performed using formulas%
\begin{equation*}
\ _{r}^{\shortmid }\mathcal{\breve{S}}:=r^{2}\partial _{r}\left( \frac{r-1}{r%
}\ _{r}^{\shortmid }\mathcal{S}\right) \mbox{ and, inversely, }\
_{r}^{\shortmid }\mathcal{S=}\frac{r}{r-1}\int_{1}^{r}dr^{\prime }\frac{\
_{r^{\prime }}^{\shortmid }\mathcal{\breve{S}}}{(r^{\prime })^{2}}.
\end{equation*}%
The implications of the inequalities for the R\'{e}nyi entropy were analyzed
for the gravitational systems with holographic description, see reviews \cite%
{solodukhin11,preskill,nishioka18}. In this subsection, the approach is
generalized for nonholonomic geometric structures and covariant mechanical
systems with applications in information theory.

\paragraph{Relative R\'{e}nyi entropy for QGIFs:}

The concept of relative entropy $\ ^{\shortmid }\mathcal{S}(\ ^{\shortmid
}\rho _{\mathcal{A}}\shortparallel \ ^{\shortmid }\sigma _{\mathcal{A}})$ (%
\ref{relativentr}) can be extended to that of relative R\'{e}nyi entropy
\cite{muller13,wilde13} (for a review, see section II.E.3b in \cite%
{nishioka18}). For a system QGIFs with two density matrices $\ ^{\shortmid
}\rho _{\mathcal{A}}$ and $\ ^{\shortmid }\sigma _{\mathcal{A}}$, we
introduce
\begin{eqnarray}
\ _{r}^{\shortmid }\mathcal{S}(\ ^{\shortmid }\rho _{\mathcal{A}}
\shortparallel \ ^{\shortmid }\sigma _{\mathcal{A}}) &=&\frac{1}{r-1}\log %
\left[ tr\left( (\ ^{\shortmid }\sigma _{\mathcal{A}})^{(1-r)/2r}\
^{\shortmid }\rho _{\mathcal{A}}(\ ^{\shortmid }\sigma _{\mathcal{A}%
})^{(1-r)/2r}\right) ^{r}\right] ,\mbox{ for }r\in (0,1)\cup (1,\infty );
\label{relatrenyi} \\
\mbox{ or }\ _{1}^{\shortmid }\mathcal{S}(\ ^{\shortmid }\rho _{\mathcal{A}}
\shortparallel \ ^{\shortmid }\sigma _{\mathcal{A}}) &=&\ ^{\shortmid }%
\mathcal{S}(\ ^{\shortmid }\rho _{\mathcal{A}}\shortparallel \ ^{\shortmid
}\sigma _{\mathcal{A}})\mbox{ and }\ _{\infty }^{\shortmid }\mathcal{S}(\
^{\shortmid }\rho _{\mathcal{A}}\shortparallel \ ^{\shortmid }\sigma _{%
\mathcal{A}})=\log ||(\ ^{\shortmid }\sigma _{\mathcal{A}})^{-1/2}\
^{\shortmid }\rho _{\mathcal{A}}(\ ^{\shortmid }\sigma _{\mathcal{A}%
})^{-1/2}||_{\infty }.  \notag
\end{eqnarray}%
Such definitions allow us to prove certain monotonic properties,
\begin{equation*}
\ _{r}^{\shortmid }\mathcal{S}(\ ^{\shortmid }\rho _{\mathcal{A}%
}\shortparallel \ ^{\shortmid }\sigma _{\mathcal{A}})\geq \ _{r}^{\shortmid }%
\mathcal{S}(tr_{s}\ ^{\shortmid }\rho _{\mathcal{A}}|tr_{s}\ ^{\shortmid
}\sigma _{\mathcal{A}})\mbox{ and }\partial _{r}[\ _{r}^{\shortmid }\mathcal{%
S}(\ ^{\shortmid }\rho _{\mathcal{A}}\shortparallel \ ^{\shortmid }\sigma _{%
\mathcal{A}})]\geq 0,
\end{equation*}%
and to reduce the relative R\'{e}nyi entropy to the R\'{e}nyi entropy using
\ a formula similar to (\ref{aux06}),
\begin{equation*}
\ _{r}^{\shortmid }\mathcal{S}(\ ^{\shortmid }\rho _{\mathcal{A}%
}\shortparallel 1_{\mathcal{A}}/k_{\mathcal{A}})=\log k_{\mathcal{A}}-\
_{r}^{\shortmid }\mathcal{S}(\mathcal{A}).
\end{equation*}

Nevertheless, the values (\ref{relatrenyi}) do not allow a naive
generalization of the concept of mutual information and interpretation as an
entanglement measure of quantum information because of possible negative
values of relative R\'{e}nyi entropy for $r\neq 1$ \cite{adesso12}. This
problem is solved by the $r$-R\'{e}nyi mutual information \cite{beigi13},
\begin{equation*}
\ _{r}^{\shortmid }\mathcal{J}(\mathcal{A},\mathcal{B}):=\min_{\ ^{\shortmid
}\sigma _{\mathcal{B}}}\ _{r}^{\shortmid }\mathcal{S}(\ ^{\shortmid }\rho _{%
\mathcal{A\cup B}}\shortparallel \ ^{\shortmid }\rho _{\mathcal{A}}\otimes \
^{\shortmid }\sigma _{\mathcal{B}})\geq 0,
\end{equation*}%
when the minimum is taken over all $\ ^{\shortmid }\sigma _{\mathcal{B}}.$
This formula reduced to the mutual information (\ref{mutualinf2}) for $r=1.$
In result, we can elaborate a self-consistent geometric-information
thermodynamic theory for QGIFs. This is possible if the statistical density $%
^{\shortmid }\rho (\beta ,\ ^{\shortmid }\mathcal{E}\ ,\ ^{\shortmid }%
\mathbf{g})$ (\ref{statedens}) is used for defining $\ ^{\shortmid }\sigma _{%
\mathcal{A}}$ (\ref{aux01}) as a probability distribution and respective von
Neumann density matrix formulation of the quantum models. It is not clear at
present if a version of relative R\'{e}nyi entropy can be elaborated for the
W-entropy.

\section{Conclusions}

\label{s5} The geometric flows of Riemannian metrics can be characterized by G. Perelman's W-entropy and associated statistical thermodynamic model with respective mean energy, mean entropy and fluctuation  parameter  \cite{perelman1}. Such constructions can be generalized for nonholonomic geometric flows (subjected to certain non-integrable, i.e. anholonomic, equivalently, nonholonomic conditions) with generalized entropy type functionals and related locally anisotropc diffusion, kinetic and thermodynamic theories \cite{vacaru00ap,vdiffusion,ruchin13}. In result, we can elaborate on advanced geometric methods for modeling relativistic geometric flows of classical and quantum mechanical systems, and modified commutative and noncommutative/ supersymmetric  gravity theories etc. \cite{ruchin13,gheorghiu16,rajpoot17}.

A series of our recent works, see \cite{vacaru19,vacaru19b} and refereces therein, is devoted to formulation and applications on the theory of geometric information flows, GIFs, and quantum information flows, QGIFs. In such approaches, the geometric thermodynamic models involve G. Perelman like entropic constructions \cite{vacaru19a} which are more general than those elaborated using the Bekenstein-Hawking surface-area entropy and respective holographic, dual CFT-gauge theory generalizations etc. \cite{bekenstein72,bekenstein73,hawking75,strominger96,faulkner14,swingle12}. New classes of generic off-diagonal solutions (various locally anisotropic cosmological ones, generalized black hole metrics) with the coefficients of metrics and generalized connections depending, in principle, on all spacetime and possible phase space coordinates can be constructed \cite{bubuianu18,bubuianu19} in general relativity and modified gravity theories. Such new classes of exact and parametric solutions, and related quantized models, are characterized by G. Perelman entropies and do not have Bekenstein-Hawking analogs.

In this article, we have focused on developing the notion of entanglement for quantum mechanical, QM, and geometric thermodynamic models derived for QGIFs. This specific problem is of utmost importance within vast domains of studies of properties of entanglement entropy of general relativistic quantum systems and, for instance, new types of QGIF teleportation, geometric flow testing, and encoding classical mechancal flow information in quantum states. In addition to the results of \cite{vjmp08,vrmp09} formulated for nonholonomic Lagrange and Hamilton variables, we elaborated such constructions for covariant classical and quantum mechanical systems and explicit applications in quantum information theory.

Finally, we note that important questions connected to entanglement of QGIF and modified gravity theories still remain as open challenges and promising research directions in modern geometric classical and quantum mechanics, thermodynamics, and modified gravity, see \cite{bubuianu19,vacaru19,vacaru19a,vacaru19b}.

\vskip3pt

\textbf{Acknowledgments:} This research develops former programs partially supported by IDEI, PN-II-ID-PCE-2011-3-0256, CERN 2012-2014, DAAD-2015, QGR 2016-2017. S. V. is grateful to D. Singleton, S. Rajpoot and P. Stavrinos for collaboration and supporting his research on geometric methods in physics.

\end{document}